




\message{Assuming 8.5" x 11" paper and .eps figures}    

\magnification=\magstep1	          
\mag \magstep1                            

\raggedbottom
\overfullrule=0pt 

\parskip=9pt

\def\singlespace{\baselineskip=12pt}      
\def\sesquispace{\baselineskip=16pt}      


 





\font\openface=msbm10 at10pt
 %

\def\Integers      {{\hbox{\openface Z}}}

\def\Reals         {{\hbox{\openface R}}}
\def\Complexes     {{\hbox{\openface C}}}

 %
 %
 %



\font\german=eufm10 at 10pt

\def\Buchstabe#1{{\hbox{\german #1}}}








%

%
%



\def\implies{\Rightarrow}

%



\def\sqr#1#2{\vcenter{
  \hrule height.#2pt 
  \hbox{\vrule width.#2pt height#1pt 
        \kern#1pt 
        \vrule width.#2pt}
  \hrule height.#2pt}}


\def\dal{\mathop{\,\sqr{7}{5}\,}}
\def\block{\dal}



\def\lto{\mathop
        {\hbox{${\lower3.8pt\hbox{$<$}}\atop{\raise0.2pt\hbox{$\sim$}}$}}}
\def\gto{\mathop
        {\hbox{${\lower3.8pt\hbox{$>$}}\atop{\raise0.2pt\hbox{$\sim$}}$}}}
%
%
%



\def\part{\subseteq}		




\def\to{\mathop\rightarrow}	

\def\orthog{\mathop\bot}

\def\ideq{\equiv}		

\def\SetOf#1#2{\left\{ #1  \,|\, #2 \right\} }

\def\less{\backslash}		


\def\interior #1 {  \buildrel\circ\over  #1}     




\def\basisvector#1#2#3{
 \lower6pt\hbox{
  ${\buildrel{\displaystyle #1}\over{\scriptscriptstyle(#2)}}$}^#3}

\def\eps{\varepsilon}

\def\alfa{\alpha}


\def\tilde{\widetilde}		
\def\bar{\overline}		
\def\hat{\widehat}		




\fontdimen16\textfont2=2.5pt
\fontdimen17\textfont2=2.5pt
\fontdimen14\textfont2=4.5pt
\fontdimen13\textfont2=4.5pt 

\let\miguu=\footnote
\def\footnote#1#2{{$\,$\parindent=9pt\baselineskip=13pt%
\miguu{#1}{#2\vskip -7truept}}}
 %

\def\linebreak{\hfil\break}
\def\lbr{\linebreak}
\def\pagebreak{\vfil\break}


\def\BulletItem #1 {\item{$\bullet$}{#1 }}
\def\bulletitem #1 {\BulletItem{#1}}

\def\REMARK{\noindent {\csmc Remark \ }}

\def\THEOREM{\noindent {\csmc Theorem \ }}

\def\LEMMA{\smallskip\noindent {\csmc Lemma }}
\def\PROOF{\noindent {\csmc Proof \ }}
\def\DEFINITION{\noindent {\csmc Definition \ }}

\def\author#1 {\medskip\centerline{\it #1}\bigskip}

\def\address#1{\centerline{\it #1}\smallskip}

\def\furtheraddress#1{\centerline{\it and}\smallskip\centerline{\it #1}\smallskip}

\def\email#1{\smallskip\centerline{\it address for email: #1}} 

\def\AbstractBegins
{
 \singlespace                                        
 \bigskip\leftskip=1.5truecm\rightskip=1.5truecm     
 \centerline{\bf Abstract}
 \smallskip
 \noindent	
 } 
\def\AbstractEnds
{
 \bigskip\leftskip=0truecm\rightskip=0truecm       
 }

\def\section #1 {\bigskip\noindent{\headingfont #1 }\par\nobreak\smallskip\noindent}

\def\subsection #1 {\medskip\noindent{\subheadfont #1 }\par\nobreak\smallskip\noindent}
 %

\def\ReferencesBegin
{
 \singlespace					   
 \vskip 0.5truein
 \centerline           {\bf References}
 \par\nobreak
 \medskip
 \noindent
 \parindent=2pt
 \parskip=6pt			
 }
 %

\def\reference{\hangindent=1pc\hangafter=1} 

\def\ref{\reference}

 %

\def\journaldata#1#2#3#4{{\it #1\/}\phantom{--}{\bf #2$\,$:} $\!$#3 (#4)}
 %

\def\eprint#1{{\tt #1}}

\def\arxiv#1{\hbox{\tt http://arXiv.org/abs/#1}}
 %


\def\webhome{{\tt http://www.perimeterinstitute.ca/personal/rsorkin/}}


\def\webtilde{\lower2pt\hbox{${\widetilde{\phantom{m}}}$}}
 %

 %

\def\hpf#1{\webhome{\tt{some.papers/}}}
 %

\def\hpfll#1{\webhome{\tt{lisp.library/}}}
 %



\font\titlefont=cmb10 scaled\magstep2 

\font\headingfont=cmb10 at 12pt
%

\font\subheadfont=cmssi10 scaled\magstep1 
%


\font\csmc=cmcsc10  





\def\THEOREM{\smallskip\noindent {\csmc Theorem \ }}





\edef\resetatcatcode{\catcode`\noexpand\@\the\catcode`\@\relax}
\ifx\miniltx\undefined\else\endinput\fi
\let\miniltx\box

\def\makeatletter{\catcode`\@11\relax}

\makeatletter

\def\@makeother#1{\catcode`#1=12\relax}

\def\@ifnextchar#1#2#3{%
  \let\reserved@d=#1%
  \def\reserved@a{#2}\def\reserved@b{#3}%
  \futurelet\@let@token\@ifnch}
\def\@ifnch{%
  \ifx\@let@token\@sptoken
    \let\reserved@c\@xifnch
  \else
    \ifx\@let@token\reserved@d
      \let\reserved@c\reserved@a
    \else
      \let\reserved@c\reserved@b
    \fi
  \fi
  \reserved@c}
\begingroup
\def\:{\global\let\@sptoken= } \:  
\def\:{\@xifnch} \expandafter\gdef\: {\futurelet\@let@token\@ifnch}
\endgroup

\def\@ifstar#1{\@ifnextchar *{\@firstoftwo{#1}}}
\long\def\@dblarg#1{\@ifnextchar[{#1}{\@xdblarg{#1}}}
\long\def\@xdblarg#1#2{#1[{#2}]{#2}}

\long\def \@gobble #1{}
\long\def \@gobbletwo #1#2{}
\long\def \@gobblefour #1#2#3#4{}
\long\def\@firstofone#1{#1}
\long\def\@firstoftwo#1#2{#1}
\long\def\@secondoftwo#1#2{#2}

\def\NeedsTeXFormat#1{\@ifnextchar[\@needsf@rmat\relax}
\def\@needsf@rmat[#1]{}
\def\ProvidesPackage#1{\@ifnextchar[%
    {\@pr@videpackage{#1}}{\@pr@videpackage#1[]}}
\def\@pr@videpackage#1[#2]{\wlog{#1: #2}}

\let\DeclareOption\@gobbletwo
\def\ProcessOptions{\@ifstar\relax\relax}

\def\RequirePackage{%
  \@fileswithoptions\@pkgextension}
\def\@fileswithoptions#1{%
  \@ifnextchar[
    {\@fileswith@ptions#1}%
    {\@fileswith@ptions#1[]}}
\def\@fileswith@ptions#1[#2]#3{%
  \@ifnextchar[
  {\@fileswith@pti@ns#1[#2]#3}%
  {\@fileswith@pti@ns#1[#2]#3[]}}

\def\@fileswith@pti@ns#1[#2]#3[#4]{%
    \def\reserved@b##1,{%
      \ifx\@nil##1\relax\else
        \ifx\relax##1\relax\else
         \noexpand\@onefilewithoptions##1[#2][#4]\noexpand\@pkgextension
        \fi
        \expandafter\reserved@b
      \fi}%
      \edef\reserved@a{\zap@space#3 \@empty}%
      \edef\reserved@a{\expandafter\reserved@b\reserved@a,\@nil,}%
  \reserved@a}

\def\zap@space#1 #2{%
  #1%
  \ifx#2\@empty\else\expandafter\zap@space\fi
  #2}

\let\@empty\empty
\def\@pkgextension{sty}

\def\@onefilewithoptions#1[#2][#3]#4{%
  \input #1.#4 }

\def\typein{%
  \let\@typein\relax
  \@testopt\@xtypein\@typein}
\def\@xtypein[#1]#2{%
  \message{#2}%
  \advance\endlinechar\@M
  \read\@inputcheck to#1%
  \advance\endlinechar-\@M
  \@typein}
\def\@namedef#1{\expandafter\def\csname #1\endcsname}
\def\@nameuse#1{\csname #1\endcsname}
\def\@cons#1#2{\begingroup\let\@elt\relax\xdef#1{#1\@elt #2}\endgroup}
\def\@car#1#2\@nil{#1}
\def\@cdr#1#2\@nil{#2}
\def\@carcube#1#2#3#4\@nil{#1#2#3}
\def\@preamblecmds{}

\def\@star@or@long#1{%
  \@ifstar
   {\let\l@ngrel@x\relax#1}%
   {\let\l@ngrel@x\long#1}}

\let\l@ngrel@x\relax
\def\newcommand{\@star@or@long\new@command}
\def\new@command#1{%
  \@testopt{\@newcommand#1}0}
\def\@newcommand#1[#2]{%
  \@ifnextchar [{\@xargdef#1[#2]}%
                {\@argdef#1[#2]}}
\long\def\@argdef#1[#2]#3{%
   \@ifdefinable #1{\@yargdef#1\@ne{#2}{#3}}}
\long\def\@xargdef#1[#2][#3]#4{%
  \@ifdefinable#1{%
     \expandafter\def\expandafter#1\expandafter{%
          \expandafter
          \@protected@testopt
          \expandafter
          #1%
          \csname\string#1\expandafter\endcsname
          {#3}}%
       \expandafter\@yargdef
          \csname\string#1\endcsname
           \tw@
           {#2}%
           {#4}}}
\def\@testopt#1#2{%
  \@ifnextchar[{#1}{#1[#2]}}
\def\@protected@testopt#1{
  \ifx\protect\@typeset@protect
    \expandafter\@testopt
  \else
    \@x@protect#1%
  \fi}
\long\def\@yargdef#1#2#3{%
  \@tempcnta#3\relax
  \advance \@tempcnta \@ne
  \let\@hash@\relax
  \edef\reserved@a{\ifx#2\tw@ [\@hash@1]\fi}%
  \@tempcntb #2%
  \@whilenum\@tempcntb <\@tempcnta
     \do{%
         \edef\reserved@a{\reserved@a\@hash@\the\@tempcntb}%
         \advance\@tempcntb \@ne}%
  \let\@hash@##%
  \l@ngrel@x\expandafter\def\expandafter#1\reserved@a}
\long\def\@reargdef#1[#2]#3{%
  \@yargdef#1\@ne{#2}{#3}}
\def\renewcommand{\@star@or@long\renew@command}
\def\renew@command#1{%
  {\escapechar\m@ne\xdef\@gtempa{{\string#1}}}%
  \expandafter\@ifundefined\@gtempa
     {\@latex@error{\string#1 undefined}\@ehc}%
     {}%
  \let\@ifdefinable\@rc@ifdefinable
  \new@command#1}
\long\def\@ifdefinable #1#2{%
      \edef\reserved@a{\expandafter\@gobble\string #1}%
     \@ifundefined\reserved@a
         {\edef\reserved@b{\expandafter\@carcube \reserved@a xxx\@nil}%
          \ifx \reserved@b\@qend \@notdefinable\else
            \ifx \reserved@a\@qrelax \@notdefinable\else
              #2%
            \fi
          \fi}%
         \@notdefinable}
\let\@@ifdefinable\@ifdefinable
\long\def\@rc@ifdefinable#1#2{%
  \let\@ifdefinable\@@ifdefinable
  #2}
\def\newenvironment{\@star@or@long\new@environment}
\def\new@environment#1{%
  \@testopt{\@newenva#1}0}
\def\@newenva#1[#2]{%
   \@ifnextchar [{\@newenvb#1[#2]}{\@newenv{#1}{[#2]}}}
\def\@newenvb#1[#2][#3]{\@newenv{#1}{[#2][#3]}}
\def\renewenvironment{\@star@or@long\renew@environment}
\def\renew@environment#1{%
  \@ifundefined{#1}%
     {\@latex@error{Environment #1 undefined}\@ehc
     }{}%
  \expandafter\let\csname#1\endcsname\relax
  \expandafter\let\csname end#1\endcsname\relax
  \new@environment{#1}}
\long\def\@newenv#1#2#3#4{%
  \@ifundefined{#1}%
    {\expandafter\let\csname#1\expandafter\endcsname
                         \csname end#1\endcsname}%
    \relax
  \expandafter\new@command
     \csname #1\endcsname#2{#3}%
     \l@ngrel@x\expandafter\def\csname end#1\endcsname{#4}}

\def\providecommand{\@star@or@long\provide@command}
\def\provide@command#1{%
  {\escapechar\m@ne\xdef\@gtempa{{\string#1}}}%
  \expandafter\@ifundefined\@gtempa
    {\def\reserved@a{\new@command#1}}%
    {\def\reserved@a{\renew@command\reserved@a}}%
   \reserved@a}%

\def\@ifundefined#1{%
  \expandafter\ifx\csname#1\endcsname\relax
    \expandafter\@firstoftwo
  \else
    \expandafter\@secondoftwo
  \fi}

\chardef\@xxxii=32
\mathchardef\@Mi=10001
\mathchardef\@Mii=10002
\mathchardef\@Miii=10003
\mathchardef\@Miv=10004

\newcount\@tempcnta
\newcount\@tempcntb
\newif\if@tempswa\@tempswatrue
\newdimen\@tempdima
\newdimen\@tempdimb
\newdimen\@tempdimc
\newbox\@tempboxa
\newskip\@tempskipa
\newskip\@tempskipb
\newtoks\@temptokena

\long\def\@whilenum#1\do #2{\ifnum #1\relax #2\relax\@iwhilenum{#1\relax
     #2\relax}\fi}
\long\def\@iwhilenum#1{\ifnum #1\expandafter\@iwhilenum
         \else\expandafter\@gobble\fi{#1}}
\long\def\@whiledim#1\do #2{\ifdim #1\relax#2\@iwhiledim{#1\relax#2}\fi}
\long\def\@iwhiledim#1{\ifdim #1\expandafter\@iwhiledim
        \else\expandafter\@gobble\fi{#1}}
\long\def\@whilesw#1\fi#2{#1#2\@iwhilesw{#1#2}\fi\fi}
\long\def\@iwhilesw#1\fi{#1\expandafter\@iwhilesw
         \else\@gobbletwo\fi{#1}\fi}
\def\@nnil{\@nil}
\def\@empty{}
\def\@fornoop#1\@@#2#3{}
\long\def\@for#1:=#2\do#3{%
  \expandafter\def\expandafter\@fortmp\expandafter{#2}%
  \ifx\@fortmp\@empty \else
    \expandafter\@forloop#2,\@nil,\@nil\@@#1{#3}\fi}
\long\def\@forloop#1,#2,#3\@@#4#5{\def#4{#1}\ifx #4\@nnil \else
       #5\def#4{#2}\ifx #4\@nnil \else#5\@iforloop #3\@@#4{#5}\fi\fi}
\long\def\@iforloop#1,#2\@@#3#4{\def#3{#1}\ifx #3\@nnil
       \expandafter\@fornoop \else
      #4\relax\expandafter\@iforloop\fi#2\@@#3{#4}}
\def\@tfor#1:={\@tf@r#1 }
\long\def\@tf@r#1#2\do#3{\def\@fortmp{#2}\ifx\@fortmp\space\else
    \@tforloop#2\@nil\@nil\@@#1{#3}\fi}
\long\def\@tforloop#1#2\@@#3#4{\def#3{#1}\ifx #3\@nnil
       \expandafter\@fornoop \else
      #4\relax\expandafter\@tforloop\fi#2\@@#3{#4}}
\long\def\@break@tfor#1\@@#2#3{\fi\fi}
\def\@removeelement#1#2#3{%
  \def\reserved@a##1,#1,##2\reserved@a{##1,##2\reserved@b}%
  \def\reserved@b##1,\reserved@b##2\reserved@b{%
    \ifx,##1\@empty\else##1\fi}%
  \edef#3{%
    \expandafter\reserved@b\reserved@a,#2,\reserved@b,#1,\reserved@a}}

\let\ExecuteOptions\@gobble

\def\@latex@error#1#2{%
  \errhelp{#2}\errmessage{#1}}

\bgroup\uccode`\!`\%\uppercase{\egroup
\def\@percentchar{!}}

\ifx\@@input\@undefined
 \let\@@input\input
\fi

\def\input{\@ifnextchar\bgroup\@iinput\@@input}
\def\@iinput#1{\input{#1} }

\ifx\filename@parse\@undefined
  \def\reserved@a{./}\ifx\@currdir\reserved@a
    \wlog{^^JDefining UNIX/DOS style filename parser.^^J}
    \def\filename@parse#1{%
      \let\filename@area\@empty
      \expandafter\filename@path#1/\\}
    \def\filename@path#1/#2\\{%
      \ifx\\#2\\%
         \def\reserved@a{\filename@simple#1.\\}%
      \else
         \edef\filename@area{\filename@area#1/}%
         \def\reserved@a{\filename@path#2\\}%
      \fi
      \reserved@a}
  \else\def\reserved@a{[]}\ifx\@currdir\reserved@a
    \wlog{^^JDefining VMS style filename parser.^^J}
    \def\filename@parse#1{%
      \let\filename@area\@empty
      \expandafter\filename@path#1]\\}
    \def\filename@path#1]#2\\{%
      \ifx\\#2\\%
         \def\reserved@a{\filename@simple#1.\\}%
      \else
         \edef\filename@area{\filename@area#1]}%
         \def\reserved@a{\filename@path#2\\}%
      \fi
      \reserved@a}
  \else\def\reserved@a{:}\ifx\@currdir\reserved@a
    \wlog{^^JDefining Mac style filename parser.^^J}
    \def\filename@parse#1{%
      \let\filename@area\@empty
      \expandafter\filename@path#1:\\}
    \def\filename@path#1:#2\\{%
      \ifx\\#2\\%
         \def\reserved@a{\filename@simple#1.\\}%
      \else
         \edef\filename@area{\filename@area#1:}%
         \def\reserved@a{\filename@path#2\\}%
      \fi
      \reserved@a}
  \else
    \wlog{^^JDefining generic filename parser.^^J}
    \def\filename@parse#1{%
      \let\filename@area\@empty
      \expandafter\filename@simple#1.\\}
  \fi\fi\fi
  \def\filename@simple#1.#2\\{%
    \ifx\\#2\\%
       \let\filename@ext\relax
    \else
       \edef\filename@ext{\filename@dot#2\\}%
    \fi
    \edef\filename@base{#1}}
  \def\filename@dot#1.\\{#1}
\else
  \wlog{^^J^^J%
    \noexpand\filename@parse was defined in texsys.cfg:^^J%
    \expandafter\strip@prefix\meaning\filename@parse.^^J%
    }
\fi

\long\def \IfFileExists#1#2#3{%
  \openin\@inputcheck#1 %
  \ifeof\@inputcheck
    \ifx\input@path\@undefined
      \def\reserved@a{#3}%
    \else
      \def\reserved@a{\@iffileonpath{#1}{#2}{#3}}%
    \fi
  \else
    \closein\@inputcheck
    \edef\@filef@und{#1 }%
    \def\reserved@a{#2}%
  \fi
  \reserved@a}
\long\def\@iffileonpath#1{%
  \let\reserved@a\@secondoftwo
  \expandafter\@tfor\expandafter\reserved@b\expandafter
             :\expandafter=\input@path\do{%
    \openin\@inputcheck\reserved@b#1 %
    \ifeof\@inputcheck\else
      \edef\@filef@und{\reserved@b#1 }%
      \let\reserved@a\@firstoftwo%
      \closein\@inputcheck
      \@break@tfor
    \fi}%
  \reserved@a}
\long\def \InputIfFileExists#1#2{%
  \IfFileExists{#1}%
    {#2\@addtofilelist{#1}\@@input \@filef@und}}

\chardef\@inputcheck0

\let\@addtofilelist \@gobble

\def\@defaultunits{\afterassignment\remove@to@nnil}
\def\remove@to@nnil#1\@nnil{}

\newdimen\leftmarginv
\newdimen\leftmarginvi

\newdimen\@ovxx
\newdimen\@ovyy
\newdimen\@ovdx
\newdimen\@ovdy
\newdimen\@ovro
\newdimen\@ovri
\newdimen\@xdim
\newdimen\@ydim
\newdimen\@linelen
\newdimen\@dashdim

\long\def\mbox#1{\leavevmode\hbox{#1}}

\let\@onlypreamble\@gobble

\let\protect\relax

\newdimen\fboxsep
\newdimen\fboxrule

\fboxsep = 3pt
\fboxrule = .4pt

\def\@height{height} \def\@depth{depth} \def\@width{width}
\def\@minus{minus}
\def\@plus{plus}
\def\hb@xt@{\hbox to}

\long\def\@begin@tempboxa#1#2{%
   \begingroup
     \setbox\@tempboxa#1{\color@begingroup#2\color@endgroup}%
     \def\width{\wd\@tempboxa}%
     \def\height{\ht\@tempboxa}%
     \def\depth{\dp\@tempboxa}%
     \let\totalheight\@ovri
     \totalheight\height
     \advance\totalheight\depth}
\let\@end@tempboxa\endgroup

\let\set@color\relax
\let\color@begingroup\relax
\let\color@endgroup\relax
\let\color@setgroup\relax

\let\color@hbox\relax
\let\color@vbox\relax
\let\color@endbox\relax


\begingroup
  \catcode`P=12
  \catcode`T=12
  \lowercase{
    \def\x{\def\rem@pt##1.##2PT{##1\ifnum##2>\z@.##2\fi}}}
  \expandafter\endgroup\x
\def\strip@pt{\expandafter\rem@pt\the}


\def\@input#1{%
  \IfFileExists{#1}{\@@input\@filef@und}{\message{No file #1.}}}

\def\@warning{\immediate\write16}




\def\Gin@driver{dvips.def}
\input graphicx.sty

\resetatcatcode




\resetatcatcode                 


\def\Caption#1{\vbox{

 \leftskip=1.5truecm\rightskip=1.5truecm     
 \singlespace
 \noindent #1
 \vskip .25in\leftskip=0truecm\rightskip=0truecm}
 \sesquispace}


\def\Fmoja{1}
\def\Fmbili{2}
\def\Ftatu{3}
\def\Fnne{4}
\def\Ftano{5}
\def\Fsita{6}

\def\SetOf#1#2{\left\{ #1  \,:\, #2 \right\} }

\def\H{\Buchstabe{H}}
\def\T{\Buchstabe{T}}
\def\Z{\Buchstabe{Z}}
\def\A{\Buchstabe{A}}
\def\S{\Buchstabe{S}}
\def\R{\Buchstabe{R}}
\def\X{\Buchstabe{X}}

\def\cyl{\mathop {{\rm \, cyl}} \nolimits}	
\def\stem{\mathop {{\rm \, stem}} \nolimits}	
\def\node{\mathop {{\rm \, node}} \nolimits}	
\def\Lim{\mathop {{\rm \, Lim}} \nolimits}	
\def\qed{$\block$}

\def\P{\mathop {{\hbox{\openface P}}} \nolimits}
\def\Pbar{\mathop {\bar{\hbox{\openface P}}} \nolimits}

\def\alephC{\aleph_1}

\def\oga{\omega}
\def\Oga{\Omega}
\def\Abar{\bar{A\,}}
\def\Stilde{\tilde{S}}

\def\ketvector#1{|#1\rangle}
\def\kv{\ketvector}

\def\return{R}
\def\nonreturn{R'}

\def\evenlyconvergent{evenly convergent }
\def\Evenlyconvergent{Evenly convergent }





\phantom{}




\sesquispace
\centerline{{\titlefont Toward a ``fundamental theorem of quantal measure theory''}\footnote{$^{^{\displaystyle\star}}$}%
{ To appear in a special issue of the journal, {\it Mathematical Structures in Computer Science}
  edited by  Cris Calude and Barry Cooper (Cambridge University Press).
}}

\bigskip


\singlespace			        

\author{Rafael D. Sorkin}
\address
 {Perimeter Institute, 31 Caroline Street North, Waterloo ON, N2L 2Y5 Canada}
\furtheraddress
 {Department of Physics, Syracuse University, Syracuse, NY 13244-1130, U.S.A.}
\email{rsorkin@perimeterinstitute.ca}

\AbstractBegins                              
   We address the extension problem for quantal measures of
   path-integral type, concentrating on two cases: sequential growth of
   causal sets, and a particle moving on the finite lattice
   $\Integers_n$.  In both cases the dynamics can be coded into a
   vector-valued measure $\mu$ on $\Oga$, the space of all histories.
   Initially $\mu$ is defined only on special subsets of $\Oga$ called
   cylinder-events, and one would like to extend it to a larger family
   of subsets (events) in analogy to the way this is done in the
   classical theory of stochastic processes.  Since quantally $\mu$ is
   generally not of bounded variation, a new method is required.  We
   propose a method that defines the measure of an event by means of a
   sequence of simpler events which in a suitable sense converges to the
   event whose measure one is seeking to define.  To this end, we
   introduce canonical sequences approximating certain events, and we
   propose a measure-based criterion for the convergence of such
   sequences.  Applying the method, we encounter a simple event whose
   measure is zero classically but non-zero quantally.

\AbstractEnds                                

\bigskip



\sesquispace
\vskip -10pt

\section{1.~Introduction}					
In order to define area, even for something as simple as a disk of unit
radius, one needs to invoke an extension theorem.  In a systematic
development [1] of plane-measure, one begins by
defining the measure $\mu$ of an arbitrary rectangle, and one then seeks
to extend the set-function $\mu$ unambiguously to subsets of the plane
that can be made from rectangles via countable processes of union and
complementation (these sets comprising the $\sigma$-algebra generated by
the rectangles).  The unit disk is such a subset, and (if we take it to
be open) it obviously can be built up as the disjoint union of a
countable family of rectangles.  But this can be done in an infinite
number of different ways, and one needs to know that the net area of the
rectangles is always the same, no matter which decomposition one chooses
and no matter in which order one chooses to perform the resulting sum.
The theorem that guarantees this consistency is known as the
Kolmogorov-Carath{\'e}odory extension theorem, but it might also be
called the ``fundamental theorem of classical measure theory''.  Not
only is it used  to construct Lebesgue measure, but it plays a
central role in defining stochastic processes like the Wiener
process, a mathematical model of Brownian motion that also describes
the Wick rotated 
path-integral for a non-relativistic
free particle on the line.

In this sort of application, one is dealing with a
probability-measure on a space of paths or more generally ``histories'',
and the possible values of $\mu$ are therefore positive real numbers
between $0$ and $1$.  When one seeks to define a genuine path integral
in real time however (as opposed to Wick-rotated, imaginary time), one
encounters complex amplitudes that can be arbitrarily large and of any
phase.  Once again, there are specially simple sets of paths, analogs of
the rectangles called ``cylinder sets'', from which the more general
sets of interest can be built up, but the sums that arise in this case
no longer converge absolutely.  In technical terms the complex measure
one is trying to extend is not of bounded variation, and the available
extension theorems cannot be used [2].

The problems that one faces vary, depending on context.  There
are ``ultraviolet'' problems springing from the infinite divisibility of
the paths or ``histories'' one is trying to sum over, and there are
``infrared'' problems that arise in connection with histories that are
unbounded in time.  By limiting ourselves to spatio-temporally discrete
processes we nullify the former problems, and that will be the context
of the rest of this paper, where we will encounter only discrete
histories like those that occur in a random walk.  It will thus be only
 issues of infinite time that will occupy us.

The concrete instances we will consider will be of one of two types,
which we can characterize by the kind of ``sample space'' or ``history
space'', $\Omega$, on which one builds.  The first instance arises in
the context of quantum gravity and more specifically within the causal
set programme.  There the discreteness reflects the finiteness of
Planck's constant, and the underlying physical process is a kind of
``birth'' or ``accretion'' process by means of which the causal set is
built up or ``grows''.  The corresponding sample-space of ``completed''
causal sets consists of all the countable, past-finite partial orders
$P$; and one is seeking to define a certain type of vector-valued
measure $\mu$ on it.  
(The dynamics determines $\mu$ only up to a unitary transformation.
 The object of direct physical interest is not $\mu$ itself but a
 certain scalar-valued set-function belonging to the class of strongly
 positive decoherence functionals or quantal measures on $\Omega$.
 However, any such a functional can be represented [3] as a measure
 on $\Omega$ which is valued in some Hilbert space $\H$.)
In the second type of example, the elements of $\Omega$ will be
discrete-time trajectories moving in a lattice that will be either the
integers modulo $n$ ($\Integers_n$) or just the integers as such
($\Integers$).  These examples correspond to a widely studied class of
processes known as ``quantal random walks'', but for us they will be
important primarily as simplified analogs of causal set growth
processes.  In that role, they are particularly illuminating because
their sample spaces are essentially the same ones that ``descriptive set
theory'' investigates.

How certain are we, though, that the quantal measures in these all
instances really need to be defined in a new way?  With the lattices,
the dynamical laws in question are those of the evolution
generated by a unitary operator or ``transfer matrix''.  In their
path-integral formulation, such unitary laws inevitably lead to measures of
unbounded variation [3], and the theorems of the
Kolmogorov-Carath{\'e}odory type are thus guaranteed to fail.  
In the more important, causal set case however, there remains some
doubt, especially given the anticipated breakdown of unitary evolution
in that case.  The only fully developed dynamics one has for causal sets
is that of the classical sequential growth (CSG) models, which in
themselves are not quantal in nature.  For them, the usual extension
theorems do suffice because one is dealing with a classical probability
measure [4].  But if one complexifies the parameters of a CSG model,
one obtains straightforwardly a family of quantal measures (decoherence
functionals) which are in general neither unitary nor of bounded
variation [3].  Although none of these complexified CSG dynamics is
likely to exhibit quite the type of interference required by quantum gravity,
the fact that the measures that arise are not of bounded variation
suggests that this might turn out to be a general feature of quantal
causal sets, just as it is a general feature of quantal path integrals
in other contexts.  
Nevertheless it's worth keeping in mind the possibility that the
physically appropriate quantal measures for causal set dynamics will
turn out to be $\sigma$-additive in the traditional sense.  Were that to
happen, quantum gravity would have revealed itself to be more tractable
mathematically than the nominally much ``simpler'' non-relativistic free
particle!  The problems addressed in the present paper would then be
pseudo-problems, as far as quantum gravity went.

In our current state of ignorance, however, it seems prudent not to
count on so much good fortune.  And besides, one might still like to
have a well defined path-integral for systems like the free particle,
without having to embed them in a full-blown theory of quantum gravity.
What, then, can one do when bounded variation fails?
As urged in reference [3], such a failure need not be the end of the
story, because in a concrete physical situation, the space of histories
has more structure than what is available in an arbitrary measure space.
Indeed, physicists routinely work with infinite sums and integrals that
converge only conditionally.  Typically one introduces a ``cutoff'' or
integrating factor in a manner mandated by physical considerations, in
effect doing the sums or integrals in a particular order so that their
convergence need not be absolute.
In the case of the planar disk, for example, instead of expressing it as
a disorganized sum of an infinite number of rectangles, one might think
to employ a definite sequence of approximations, each consisting only of
rectangles bigger than a certain size $\epsilon$.  The area would then
be given by the $\epsilon\to0$ limit of these approximations.

In our situation, one can attempt something similar by considering
``late-time'' cylinder sets to be ``finer'' than ``early-time'' ones.
In order to implement this idea we will seek first of all, for any given
set $A\subseteq\Omega$ of histories or paths, a ``canonical'' sequence
of approximations $A_n$ to $A$ in terms of cylinder sets, and this
sequence should be as near to unique as feasible.  Then, given such a
sequence, we will try to decide what further convergence properties it
ought to have in order that we can form a limit $\mu(A)$ of the individual
$\mu(A_n)$ and consistently attribute this limit to $A$ as its quantal
measure.  

In what follows, we take only a few steps in the direction indicated,
pointing out along the way various pitfalls that one needs to avoid.
Hopefully this can at least illustrate the kind of approach one
might take to the extension problem for quantal measures.
Sections 6 and 7 are the heart of the paper.
In section 6, we will define canonical approximations for a limited
class of {\it events}, as (sufficiently regular) subsets
$A\subseteq\Omega$ are normally designated.  
We will then introduce, in section 7, a convergence criterion 
for an approximating sequence $A_n$, and we
will prove that the resulting extension of $\mu$ is additive for
disjoint unions of open sets.  
Limited as our approximation scheme will be,
it will at least embrace the type of event $A$ which is most important for
the sake of causal sets, namely the {\it covariant stem-event}.  Among
all the sets of histories to which one might wish to assign a measure,
the only indispensable ones are these.  Without their aid it would be
nearly impossible to produce a generally covariant dynamical scheme in
any useful sense [5] [4].

An appendix lists some of the symbols used in the body of the paper.

\bigskip

\let\parskipsave=\parskip
\parskip=1pt
\singlespace
\centerline {\bf Contents}

\item{1.} Introduction

\item{2.} {Sample-spaces and amplitudes for causal sets and the 2-site hopper}
\itemitem{} causal sets
\itemitem{} {$n$-site hopper}

\item{3.} {Some events whose measures one would like to define}

\item{4.} {$\Omega$ as a compact metric space}
\itemitem{} {open and closed sets}
\itemitem{} {the tree of truncated histories}

\item{5.} {Set-theoretic limits of events}
\item{6.} {Canonical approximations for certain events}

\item{7.} {\Evenlyconvergent  sequences of events }
\itemitem{} {Examples}

\item{8.} {Epilogue: does physics need actual infinity?}
\item{} \hskip -12pt Appendix. {Some symbols used, in approximate order of appearance}

\parskip=\parskipsave
\sesquispace

\bigskip

\section{2.~Sample-spaces and amplitudes for causal sets and the 2-site hopper}
\subsection{causal sets}
A causal set (or {\it causet}) 
[6] [7] [8] [9] [10]
in its most general conception can be any
locally finite partial order or {\it poset}, but in the context of the
dynamics of sequential growth and quantal cosmology no element of the
causet will possess more than a finite number of ancestors.  For present
purposes we may thus define a causet as a {\it past-finite countable poset}, 
i.e. a countable (possibly finite) set of {\it elements} endowed with a
transitive, acyclic order-relation, $\prec$, which I will also take to
be irreflexive.  These concepts are exposed in greater detail in
[11], where the notion of sequential growth is also
explained.  Here I will just summarize the main definitions and
introduce the notation we will use.

A sequential growth process proceeds as a succession of ``births'' of new
elements, and in this sense is never ending.  If however, one idealizes
it has having ``run to completion'', it will have produced 
a {\it completed causet} as defined above: 
a countable set of elements, each having a finite number
of predecessors or {\it ancestors} but a possibly infinite number of
descendants.  The set of all such causets constitutes the natural
sample-space $\Omega$ for this process.  Actually, one must distinguish
here two distinct sample spaces, which one may call $\Omega^{gauge}$ and
$\Omega^{physical}$.  The latter, which in some sense is the true
sample-space, consists of {\it unlabeled} causets, or equivalently
isomorphism equivalence classes of causets.  The former, which I'll
normally denote simply as $\Omega$, 
then consists of the {\it naturally labeled} causets, a natural labeling
being a numbering, $0,1,2,\dots$ of the elements which is compatible with
the defining order $\prec$~: if $x\prec y$ then $y$ carries a bigger
label than $x$.  Here again, one of course really intends isomorphism
equivalence classes of labeled causets (or if you like the elements
could be taken to be the integers themselves in this case).

The labels record the order of the respective births, and what is most
important for us here is that this order is supposed to be fictitious in
the same sense as is a choice of coordinate system for a continuous
spacetime is fictitious.  The physically meaningful or {\it{}covariant}
events will thus correspond to subsets of $\Omega^{physical}$, whereas
the measure $\mu$ defining the growth process is in the first instance
defined on $\Omega^{gauge}$.
But even a very simple subset of $\Omega^{physical}$, even a
singleton, will equate to a much less accessible subset
of $\Omega^{gauge}$, namely the subset obtained by taking every possible
natural labeling of every member of the original subset.  Thus arises
the need for an extension of $\mu$ that will assign well defined
measures to such ``covariant'' subsets of $\Omega^{gauge}$.  Unlike for
the example of the hopper to be discussed next, this is not just
a matter of convenience if one wants to be in a position to ask truly
label-independent questions about the causet. 

Henceforth in this paper all causets will be labeled unless otherwise
specified.
(In reference [4] the true or ``covariant'' sample space
 was denoted simply by $\Omega$, while its labeled counterpart was
 $\tilde\Omega$.  Here however, it seems simpler to use $\Omega$ for the
 latter, since it is the space we will usually be dealing with.)

In reference [4], the measures defining the CSG
dynamical models were defined rigorously by extending a probability
measure given originally on the space $\Z$ of {\it cylinder events} (or
cylinder sets), where a cylinder event $\cyl(c)\in\Z$ is by
definition the set of all completed causets containing a given,
naturally labeled, finite causet $c$.  A finite causet will also be
called a {\it stem} and on occasion a ``truncated history''.  
In conjunction with these definitions, let us also define $\Omega(n)$,
the space of all naturally labeled causets of $n$ elements, 
and $\Z(n)$ or $\Z_n$, the space of cylinder events of the form
$\cyl(c)$ for $c\in\Omega(n)$.
The cylinder sets
comprise what is called a ``semiring'' of sets in the sense that given
any two cylinder sets, $Z_1$ and $Z_2$, their intersection,
$Z_1{}Z_2 \ideq Z_1{\cap}Z_2$, is also a cylinder event, and their
difference $Z_1\less Z_2$ is the disjoint union of a finite number of
cylinder events.
(In fact the cylinder events form an especially simple kind of semiring,
because any two of them are either disjoint or nested.)

To rehearse the definition of the CSG models in general would take
us too far afield, but the special case of ``complex percolation'' is
simple enough to be given here in illustration of the general scheme.
The vector measure $\mu$ 
is determined in this case
by a single complex parameter $p$, and it takes its
values in a one-dimensional Hilbert
space that we may identify with $\Complexes$, so that $\mu(A)$ is itself
just a complex number.
Now let $c\in\Omega(n)$ be a labeled causet of $n$
elements  and let $Z=\cyl(c)$ be the corresponding cylinder set.
Then $\mu(Z)=p^L(1-p)^I$, where
$L=L(c)$ is the number of links in $c$ and
$I=I(c)$ is the number of incomparabilities.  
Here an incomparability is simply a pair of unrelated elements, 
and a link is a causal relation, $x\prec y$, 
which is ``nearest neighbor'' 
in the sense that
there exists no intervening $z$ for which $x\prec z\prec y$.
%

Observe 
now that 
the collection of naturally labeled finite causets, i.e.
the space $\bigcup_n\Omega(n)$,
has itself the structure
of a poset in a natural way.  Indeed this poset is actually a tree $\T$,
because  its elements are labeled.  (The corresponding
structure formed by the unlabeled stems is a more interesting
poset called {\it poscau} in reference [11].)  Clearly, a
particular realization of the growth process, or equivalently the
resulting completed causet in $\Omega$, can be conceived of as an upward
path through this tree.  An analogous conception will be possible for
the two site hopper, and in this guise seems to be exploited heavily in
descriptive set theory [12][13].
(See figures \Fmoja {} and  \Fmbili.)

\vbox{
   \bigskip

  \includegraphics[scale=0.4]{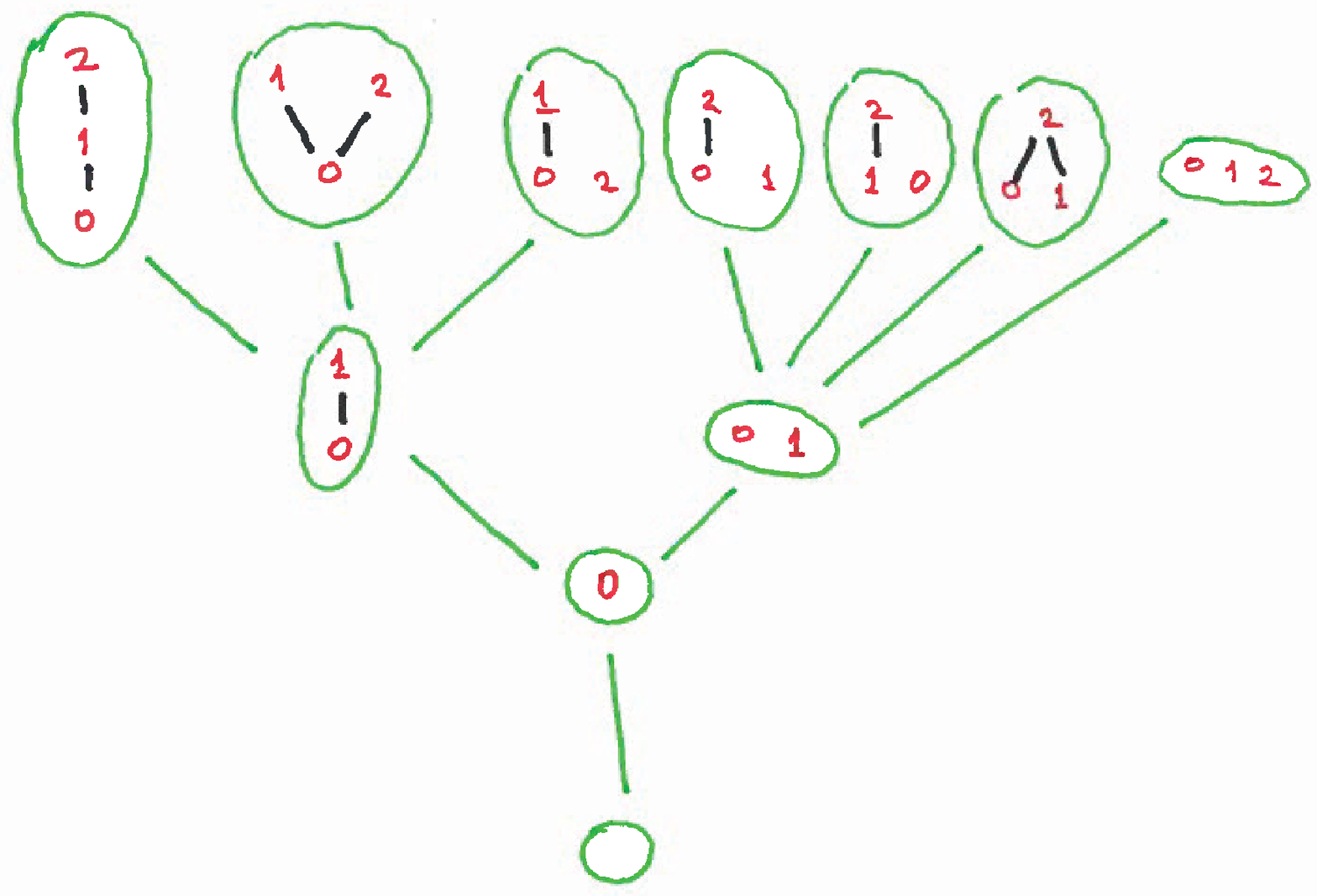}

  \Caption{{\it Figure \Fmoja.} The first 4 levels of the tree $\T$ of naturally labeled causets }}

Finally, let us define an {\it event algebra} to be a family of subsets
of the sample space $\Oga$ closed under the operations of intersection and
complementation.  An event algebra is thus a Boolean algebra or ``ring
of sets''.  To the extent it can be achieved, one normally wants the
domain of $\mu$ to be such an algebra, because for example, if the
events ``$A$ happens'' and ``$B$ happens'' are of interest, then so also
is the event ``either $A$ or $B$ happens''.  The cylinder sets $\Z$ do
not themselves form an algebra, but the family $\S$ of finite
unions of cylinder sets does.  
It is in fact $\R(\Z)$ the Boolean algebra generated by $\Z$.
In all cases of interest $\mu$ will
automatically extend uniquely from $\Z$ to $\S$, yielding a
finitely-additive measure thereon.  The space $\S$ thus constitutes a
minimum domain of definition for the vector-measure $\mu$.  The question
then will be how far $\mu$ can be extended beyond $\S$ into the
$\sigma$-algebra generated thereby, the hope being that the enlarged
domain 
$\A$
will itself be an event algebra, and that it will contain enough
events so that, at a minimum, the physically most important questions
will become well posed.  (Some noteworthy instances of covariant
questions/events will be discussed in the next section.)

\subsection{$n$-site hopper}
By ``{2-site hopper}'' I mean the formalization of a particle residing
on a 2-site lattice and at each of a discrete succession of moments
either staying where it is or jumping to the other site [14].  For
definiteness, I will assume that the moments are labeled by the natural
numbers, the sites by $\Integers_2$, and that at moment $0$ the hopper
begins at site $0$.  The definitions of sample space, cylinder event,
etc. are closely analogous to those given above for causets, and
references to them should be understandable without their formal
definitions, which I will postpone until after the transition amplitudes
have been specified.  The full course of the motion, idealized as having
run to completion, will be called a {\it path} or ``history''.  Notice
that, modulo the small ambiguity in how a real number can be expressed
as a ``binary decimal'', each such path can be identified uniquely with
a point in the unit interval $[0,1]\subseteq\Reals$.

Aside from a simpler sample space than in the causet case, the hopper
offers us in addition a fuller illustration of the problems of defining
the vector-measure corresponding to a path integral.  Unlike the former
case, where the correct choice of quantal amplitudes is only
conjectural, there exists for the hopper a choice that can be
interpreted as a straightforward discretization of the Schr{\"o}dinger
dynamics of a non-relativistic free particle moving on a circle
(cf. [15]).

These amplitudes can be understood more easily if one sets them up, not
just for two sites, but for the more general case of 
the circular lattice $\Integers_n$ (``$n$-site hopper'').  Perhaps they
will look most familiar if presented as the unitary evolution operator
or ``transfer matrix'' analogous to the propagator that solves the
Schr{\"o}dinger equation in the continuous case.  To that end, 
let $x\in\Integers_n$ be the location of the particle at some moment $t$,
let $x'$ be its location at the next moment $t'=t+1$, 
and write for brevity $\exp(2\pi i z) \ideq {\bf 1}^{z}$.
The amplitude to go from $x$ to $x'$ in a single step 
is then
$$
     {1 \over \sqrt{n}} \ {\bf 1}^{ (x-x')^2 / n}
$$
for $n$ odd and 
$$
    {1 \over \sqrt{n}} \ {\bf 1}^{(x-x')^2 / 2n}
$$
for $n$ even.  For example, for $n=6$ and with $q={\bf 1}^{1/12}$, the 
(un-normalized) amplitudes to hop by 0, 1, 2 or 3 sites are respectively
$q^0=1$, 
$q^1=q$, 
$q^4$, and
$q^9=-i$. 
For the 2- and 3-site hoppers, the above amplitudes are particularly simple,
yielding for $n=3$ the transfer matrix
$$
    {1 \over \sqrt{3}} \pmatrix{1 & \omega & \omega \cr 
                               \omega & 1 & \omega \cr 
                               \omega & \omega & 1 \cr } 
    \qquad\qquad (\omega=1^{1/3})
$$
and for $n=2$ the transfer matrix
$$
             {1 \over \sqrt{2}} \pmatrix{1 & i \cr i & 1} \ .   \eqno(1) 
$$

From these expressions and the definition of the decoherence functional it
is not hard to carry out the construction of the equivalent vector measure
along the lines of [3].  
In the simplest case of two sites,
which will be our main example herein, 
$\mu$ is valued in a two-dimensional
Hilbert space $\Complexes^2$ and, 
with a convenient choice of basis vectors, 
can be expressed as follows.  
Let $(0 \, x_1 \, x_2 \, x_3 \dots x_m)$ be a truncated path and let
$Z\subseteq\Omega$ be the corresponding cylinder event.  
Then $\mu(Z)\ideq\ketvector{Z}$
will be the two-component complex vector $v_\alfa$ where
(no summation implied)\footnote{$^\star$}
{Another notation for $v_\alfa$ could be 
 $\langle\alfa\ketvector{0 \, x_1 \, x_2 \, x_3 \dots x_m}$}
$$
    v_\alfa = 
 (U^{-m})_{\alfa x_m} U_{x_m x_{m-1}}\cdots U_{x_3 x_2} U_{x_2 x_1} U_{x_1 0} \ ,
 \eqno(2)
$$
$U$ being the unitary matrix of equation (1).
Notice incidentally that $U^{j}$ 
is periodic with period 8 
and is very easy to compute explicitly, since
$U^4=-1$ while $U^2=\pmatrix{0 & i \cr i & 0}$ is also very simple.

Finally the formal definitions for the 2-site hopper.  A truncated
history, the counterpart of a finite causal set, is for the hopper an
initial segment of a path, for example $(0,1,1,0,1)$.  (Recall our
boundary condition that all paths begin at zero.)
The set of all such truncated histories which have length $n$ when the initial
$0$ is omitted will be $\Omega(n)$, and the corresponding cylinder
events will be the elements of $\Z_n$.  The semiring $\Z$ will be the union of the
$\Z_n$.  For example, $\cyl(0,1,1,0,1)\in\Z_4$ is the set of all
completed paths of the form $(0,1,1,0,1,x_5,x_6,\cdots)$.  Exactly as above,
$\S$ will be the Boolean algebra generated by  $\Z$.  One
can check straightforwardly that $\mu$,
as defined by (2),
extends uniquely and consistently
to each $\S_n$ and therefore to $\S$ as a whole.  
Again, the truncated histories 
can be construed as the nodes of a tree $\T$, the
``branches'' or ``edges'' being given by extension of path.  
(See Figure \Fmbili.)
For example, there will be an edge from $(0,1,1,0)$ to $(0,1,1,0,1)$.

\vbox{

  \bigskip
  \includegraphics[scale=0.4]{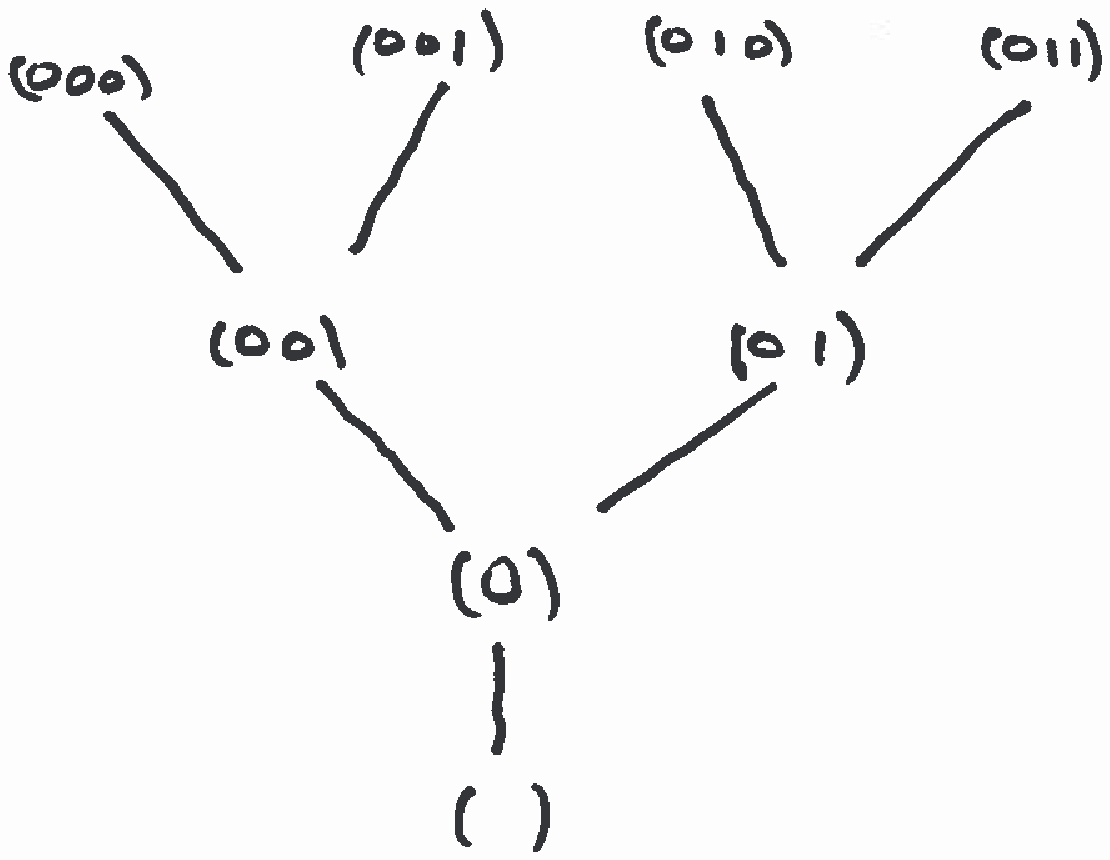}

  \Caption{{\it Figure \Fmbili.} The first 4 levels of the tree $\T$ for the 2-site hopper }}

In what follows, it will sometimes be enlightening to consider
hopper-paths on the infinite lattice $\Integers$.  In that case the
paths will be restricted to move no more than one site per step
(``random walk''), in order that the resulting tree $\T$ continue to
have a finite number of branches emanating from each node.

For an extensive discussion of the quantal 2-site hopper see
[14].  For more general sorts of quantal random walks see
[16].

\section{3.~Some events whose measures one would like to define}
The event algebra $\S$ generated by the cylinder events supplies enough
events to allow one to ask any question\footnote{$^\dagger$}
{The words ``event'' and ``question'' are in a certain sense synonyms.
 To an event $A\part\Oga$ corresponds the question ``Does $A$ happen?''.
 Note in this connection that (except in the classical case) 
 it would lead to confusion if one read ``$A$ happens'' 
 as ``the path is an element of $A$'',
 cf. [17].}
about the process under consideration, as long as it doesn't refer to
happenings arbitrarily far into the future.  But often one does not want
to be bound by this limitation, especially since in the causet case, the
``time'' referred to contains a large element of gauge, as explained
above.

To illustrate how ``infinite-time'' events enter the story, let
us dwell on a few examples, beginning with the $n$-site hopper.
%
Perhaps the simplest and most familiar 
example
of this kind is the event
$R$ of {\it return}, which occurs if and when the particle returns to
its starting point at some later time.  This event, in other words, is
the set of all paths $(0\, x_1\, x_2\, \dots)$ for which one of the
$x_i=0$.  Plainly $R$ is not in $\S$, because the return, although it
must occur at a finite time if it occurs at all, can take place
arbitrarily late.
%
For a {\it classical} hopper
on a finite lattice, 
one knows that $\mu(R)$,
the measure of the return-event (which classically is its probability),
is unity, but to express this fact directly, we need $R$ to be in the
domain of $\mu$.  Of course, we could circumvent any direct reference to
$R$ by introducing the finite-time event $R_n$ that the particle returns
on or before the $n^{th}$ step.  Instead of asserting that $\mu(R)=1$,
we could then say ``The sequence $\mu(R_n)$ converges to $1$ as
$n\to\infty$''.  Plainly, the first formulation is simpler and less
cumbersome to work with.  
Notice in this case that not only at the level of the measures, but even
at the level of the events themselves, $R$ is the limit of the $R_n$ in
a natural sense, since the latter 
are {\it nested} and
``increase monotonically to $R$''.
That is,
one has
~$R_1\part R_2\part R_3\cdots$~, 
with $R$ itself being the union of the $R_n$,
or logically speaking their ``disjunction''.
%
%
%
Were $\mu$ a classical measure, this would guarantee 
convergence of the $\mu(R_n)$ and 
consistent extension of the domain of $\mu$ 
to include the event $R$; 
in the quantal case it guarantees nothing.

A similar event to ``return'', but one which is related even less
directly to any cylinder event, is the event $R^\infty$ that the
particle visits $x=0$ infinitely often.  This event also has a
well-defined probability of unity in the classical case.  Since,
however, it cannot come to fruition at any finite time, it cannot ---
unlike the event $R$ of simple return --- be expressed as a union of
cylinder sets or other members of $\S$.  
Instead it is a countable intersection of 
events, each of which is 
a
countable union
of 
events
in $\S$.
For example, let $E(j,k)$ for $j<k$ be the event that $x_k=0$.  Then
$R^\infty = \bigcap_j \bigcup_k E(j,k)$.  
(In words: for each moment $j$
there is a later moment $k$ at which the particle visits the origin.\footnote{$^\flat$}%
{One can often arrive at such combinations by beginning with a formal
 statement of what it means for the event to happen.  In this case one
 might first write down what it means for $R^\infty$ {\it not} to occur:
 $(\exists n_0)(\forall n>n_0)(x_n\not=0)$, and then negate it to obtain
 $(\forall n_0)(\exists n>n_0)(x_n=0)$.  The nested combination of
 unions and intersections is basically just a translation of this second
 statement into set-theoretic language.})
To give meaning to $\mu(R^\infty)$ by prolonging the initially
defined measure with domain $\S$ one would thus have to think in terms
of a limit of limits.

As a third example (restricted this time to one of the lattices, 
$\Integers$ or $\Integers_n$ with $n>4$),
consider the event that the particle visits $x=3$ but never reaches
$x=5$.   Intermediate between the two previous examples in its
remoteness from $\S$, this event is naturally expressed as the
set-theoretic difference of two limits of finite-time events, the first
being, naturally the event $F$ that the particle reaches $x=3$ and the
second $G$ that it reaches $x=5$.  Just as with the return event $R$,
the event $F\less G$ is, in a well defined sense to which we will return below, a
limit of events in $\S$, but it is not simply the union or intersection
of a monotonically increasing or decreasing sequence.

It is useful at this point to introduce some further notation to help in
discussing the types of events we have just met.  
Let $\X$ be any collection of subsets of $\Oga$ closed under pairwise
union and intersection.  Then $\bigvee\X$ will be the family of events
of the form $\bigcup_{n=1}^\infty X_n$~, where $X_n\in\X$ and 
$X_1\part X_2\part X_3\cdots$~.
It is easy to see that $\bigvee\X$ is also closed under union and
intersection, also that it would not change if we dropped the
monotonicity condition, $X_1\part X_2\part X_3\cdots$~.
In words,
the members of $\bigvee\X$ are 
the unions of monotonically increasing events in $\X$.  
For the intersections of monotonically decreasing events in $\X$, I
will write dually $\bigwedge\X$.  
And for the Boolean algebra generated by $\X$, 
I will write, as above, $\R \, \X$ or $\R(\X)$~.  
Our first example, ``return'',
is then an element of $\bigvee\S$, our second of $\bigwedge\bigvee\S$,
and our third of $\R\bigvee\S$, while for $\S$ itself, we have
$\S=\R(\Z)$.

Turning now to events for causal sets, we will encounter some types
very similar to those just discussed.  Foremost in importance are the
unlabeled stem-events mentioned earlier.  Given two causets $c$  and
$c'$, of which the first is finite,
we say that $c'$ admits $c$ as a {\it stem} (or ``partial\footnote{$^\star$}
{One can also define ``full stems'' [11], but there is
 no special reason to consider them here.}
stem'') if
$c'$ contains a downward-closed subset that is isomorphic to $c$.  In
the context of sequential growth, this can also be expressed by saying
that it might have happened that elements of $c$ were all born before any of
the remaining elements of $c'$.   A stem thus generalizes the notion of
``initial segment''.   The {\it stem-event} `$\stem(c)$' is then the set of all
$c'\in\Omega$ which admit $c$ as a stem.  The stem $c$ that enters this
definition is taken to be
unlabeled, because our purpose is to produce a
label-independent or ``covariant'' event.  It is evident that $\stem(c)$
is indeed covariant in this sense, since the condition that defines it
does not refer to the labeling of $c'$.  

The importance of the stem events physically is that essentially any
covariant question that we care to ask about the causet can in principle
be phrased in terms of stem-events.
The precise result proven in [4] is that every covariant
event is equal, up to a set of measure zero, to a member of the
$\sigma$-algebra generated by the stem-events.
One can also prove that any covariant event which is open in the
topology of section 4 below is a countable union of stem events, a
purely topological result that holds independently of any assumption
about the measure $\mu$.
Ideally then, the domain of $\mu$ would embrace the whole
$\sigma$-algebra generated by the stem-events.  At a minimum, one would
hope that it would embrace the stem-events themselves.

Now
the event `$\stem(c)$' does not belong to
the domain $\S$ on which $\mu$ is initially defined, because it is not a
finite-time event when referred to ``label-time''.  If it were, then
there would exist some integer $N$ such that if the growing causet $c'$
admitted $c$ as a stem, it would already admit it as soon as the first
$N$ elements had been born.  But in fact there is nothing in principle
to stop the stem in question appearing at an arbitrarily late stage of
the growth process.  Evidently, the situation is like that of the hopper
event ``return''.  Based on this analogy, one would expect the stem
events to be found in $\bigvee\S$, and so they are, as follows directly
from the fact that any stem-event is a union of cylinder sets:
$$
   \stem(c) = 
    \bigcup 
    \SetOf 
        {\cyl(\,\tilde{b}\,)\in\Z}
        {\tilde{b} \hbox{ admits } c \hbox{ as a stem }}  \ .
  \eqno(3)
$$
The problem of extending the vector-measure $\mu$ from $\S$ to
$\bigvee\S$ is thus the most basic one for causal sets.

Starting from the stem-events, one can build up other covariant events,
whose occurrence or non-occurrence is of interest for cosmology.  The
simplest of these is the event that the causet is ``originary'', meaning
that all its elements descend from a unique minimal element or
``origin''.  To say that a completed causet is originary is simply to
say that it contains no second minimal element, for which it is
necessary and sufficient that it fail to admit the 2-element antichain
as a stem. (An antichain is a set of elements which are mutually
unrelated or ``spacelike'' to one another.)  
Thus the event `originary' is
the complement of the event `$\stem(a)$', where $a$ is the antichain
of two elements.  As such it belongs to $\bigwedge\S$, since as we have
seen, the stem events all belong to $\bigvee\S$, and union turns into
intersection under complementation.

If an originary causet represents a certain kind of ``big bang'' then a
causal set containing what is called in the combinatorics literature a
{\it post} describes a ``cosmic bounce''.  (A post is an element of a
poset which is spacelike to no element.)  In its degree of remoteness
from the elementary cylinder events, the post event is comparable to the
event of ``infinite return'' in the case of a random walk, the
similarity being even closer if we compare the post event to the
{\it{}complement} of the infinite return event.  In fact both events
belong to $\bigvee\bigwedge\S$, although this is less easy to
demonstrate for the post-event than it is in the case of return.
To see why it is nevertheless true, imagine watching a succession of
births of causet elements, $x_0, x_1, x_2\dots$ and waiting for a post
to be born.  If the birth in question is that of element $x_n$ then
$x_n$ must have every previous element as an ancestor: $x_j\prec x_n$
for all $j<n$.  This renders $x_n$ momentarily a 
``candidate for becoming a post'', 
but it does not guarantee that $x_n$ will remain a
viable candidate forever.  In order for that to occur, every subsequent
element, $x_{n+1}, x_{n+2},\cdots$, must arise as a descendant of $x_n$,
i.e.  $x_j\succ x_n$ for all $j>n$.
By thinking of the post event $P$ in this manner, namely as the set of
all sequences of births satisfying the condition that a candidate post
appear at some stage $n$ and then not lose its viability at any later
stage $m>n$, one can deduce that $P$ belongs to $\bigvee\bigwedge\S$.
The most ``covariant'' (albeit not the most direct) construction along
these lines proceeds by first expressing $P$ in terms of stem-events;
this will also illustrate the thesis that all covariant questions of
interest can be expressed in terms of stem-events.

Proceeding in this manner, note first that if $x$ is a post then its
exclusive past $T=\SetOf{y}{y\prec{x}}$ is not only a stem, but what has
been called a ``turtle'' [18], meaning in the present
context a stem that wholly precedes its complement:  
$(\forall x\in T)(\forall y\notin T)(x\prec y)$.  
Some thought reveals that a causet contains a turtle of $n$ elements iff
every stem of cardinality $n+1$ has a unique maximal element.
Introducing the term {\it principal} for such a stem, together with the
terms $n$-stem [resp. $n$-turtle] for a stem [turtle] of $n$ elements,
we can say succinctly that
{\it a causet contains an $n$-turtle iff every $(n+1)$-stem is principal}.
Furthermore it's easy to demonstrate that $x$ is a post iff both its
exclusive {\it and} inclusive pasts are turtles (the exclusive past
being $\SetOf{y\not=x}{y\prec x}$ and the inclusive past being
$\SetOf{y}{y\preceq x}$).  Therefore, in a labeled causet, element $x_n$
is a post iff every stem of either $n+1$ or $n+2$ elements is principal.
Let $P_n$ be the event that this happens, then the post event itself is
$P=\cup_{n=1}^\infty P_n$.  

Now let us examine the event $P_n$ more closely.  It {\it fails} to
happen iff some $(n+1)$-stem or $(n+2)$-stem fails to be principal.  Let
$S^n_1, S^n_2,\dots, S^n_{K_n}$ be an enumeration of all such stems (there
being only a finite number of $n$-stems, for any $n$), and let
$Q^n_j=\stem(S^n_j)$ be the corresponding stem-events.  
We then obtain $P_n$ in the ``manifestly covariant'' form 
$P_n=\Omega\less(\cup_j Q^n_j)=\Omega\less(\cup_j \stem(S^n_j))$.
$P$ is thus a countable union of finite Boolean combinations of stem
events:
$$
  P = \bigcup\limits_{n=0}^\infty \, 
      ( \Omega \, \less \, 
      (\, \bigcup\limits_{j=1}^{K_n} \stem(S^n_j)) ) \ . 
  \eqno(4)
$$
%
If one knew how to take stem events as primitive,  $P$ would thus be a rather
simple type of event,
inasmuch as the inner union only ranges over a finite number of events.
But given that the extant dynamical schemes all
begin with labeled causets, we will still need to trace everything back to the
cylinder events $\Z$.

First, though, a simple example might be in order, say for $n=1$.  
(The event $P_0$ is just the originary event, which one might not even
want to count as a post.)
The event $P_1$ requires that all 2- and 3-stems be principal.  
The only 2-stem that can occur is thus the 2-chain ($a\prec b$), while
the admissible 3-stems are the 3-chain ($a\prec b\prec c$) and the
``$\Lambda$-order'' ($a\prec c, b\prec c$).  
The stems that must be excluded --- those denoted above by $S^n_j$ ---
are correspondingly the 2- and 3-stems which are not principal: the
2-antichain, the 3-antichain, the ``L-order'' ($a\prec b, c$), and the
``V-order'' ($a\prec b, a\prec c$).
(See figure \Ftatu.)

\vbox{

  \bigskip
  \includegraphics[scale=0.4]{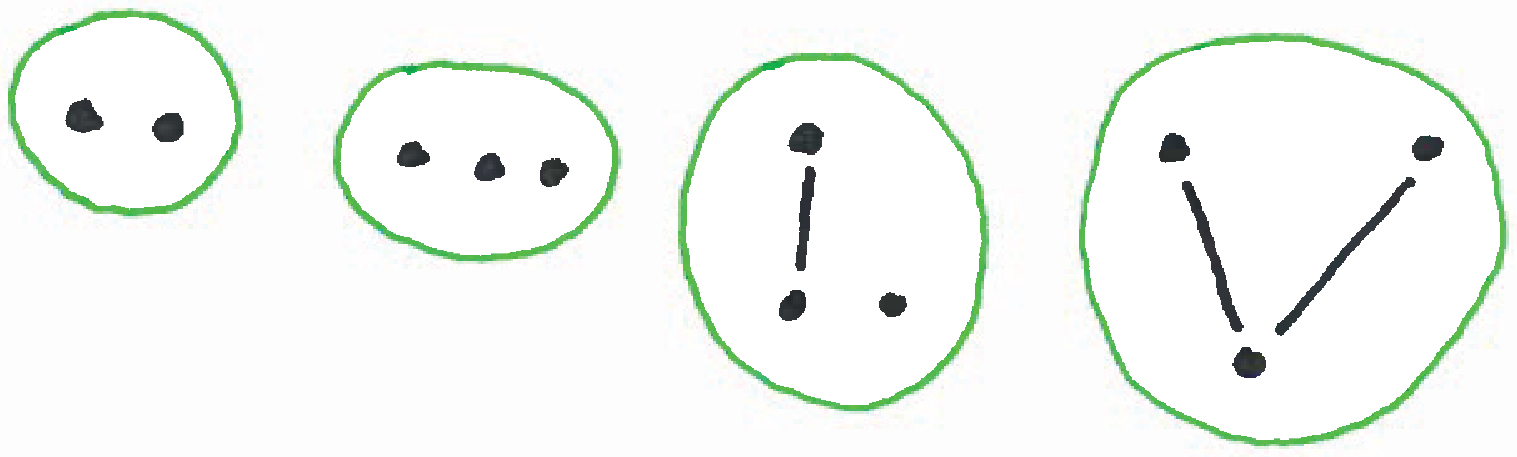}

  \Caption{{\it Figure \Ftatu.} The non-principal 2- and 3-stems. }}

To complete the demonstration that $P\in\bigvee\bigwedge\S$, let us
return exclusively to labeled causets, observing first that 
in view of equation (4), it suffices to show that the
complement of a finite union of stem-events belongs to $\bigwedge\S$. 
(Strictly speaking, given how we have defined the operation $\bigvee$,
we also need to convert the outer union in (4) into an
{\it{}increasing} countable union of events in $\bigwedge\S$.  That this
is possible follows readily from the relation (6)
of Section 5, which informs us that, when $A_n$ and $B_n$ are
both decreasing sequences of sets, then the union of their limits
coincides with the limit of the decreasing sequence
$A_n{\cup}B_n$, in consequence of which
$\bigwedge\S$ is closed under finite union, 
and
we can replace a countable
union $\cup_n F_n$ of events  $F_n\in\bigwedge\S$ with the increasing union
$\cup_n F'_n$, where $F'_n=\cup_{m\le n}F_n$~.)
To that end, recall that any stem-event $A$ is an increasing union of
events in $\S$.  Its complement, $\Omega\less A$, is therefore a
decreasing intersection of complements of events in $\S$, each of which
is itself in $\S$ since the latter, being a Boolean algebra, is closed
under complementation.  Hence, the complement of a stem-event belongs to
$\bigwedge\S$, and the same holds for the complement of a finite union
of stem-events, such as occurs in (4).  

For reasons that will become clear shortly, it is natural to designate
the elements of $\S$ as {\it clopen}, meaning ``both closed and open''
in the sense of point-set topology.\footnote{$^\dagger$}
{In the context of abstract measure theory, the term ``elementary sets''
 was used in reference [1] to refer to events
 analogous to those of $\S$.}
The events in $\bigvee\S$ will then be {\it open}, those of
$\bigwedge\S$ will be {\it closed}, and we will have expressed our
post-event $P\subseteq\Omega$ as an increasing limit of closed subsets
of $\Omega$.
Continuing in this vein, more elaborate combinations of the clopen
events can be formed, including for example the event that infinitely
many posts occur.  But the physical relevance of such combinations seems
to shrink rapidly as their complexity grows.  
Indeed, one might feel that, questions of convenience aside, no event
more complicated than a finite Boolean combination of stem-events can
claim to be indispensable.
One might even go farther and call into doubt the status of
complementation (negation), leaving unquestioned only those events
formed as finite unions and intersections of stem-events.

\section{4.~$\Omega$ as a compact metric space}                 
\subsection {open and closed sets}
Whenever the idea of convergence plays a role, one can expect, almost by
definition, that topology will make an appearance.  In the present
situation, we are talking about convergence to a given event
$A\subseteq\Omega$ of a sequence of approximating events $A_n$, in the
first instance events formed as finite unions of cylinder events and
thus belonging to the event-algebra $\S$.  In setting up such a sequence
of approximations, we would like, as explained earlier, to regard those
cylinder events that specify a greater portion of the history as more
``fine grained'' than those that specify a lesser portion.  This leads
very naturally to a definition of distance between histories that makes
$\Omega$ into a compact metric space [4] [12]. 

As implemented for causal set growth processes, the definition runs as follows.
For each pair of completed labeled causets ${a}$, ${b}\in\Omega$, we set:
$$
    d({a},{b})=1/2^n \ ,           \eqno(5)  
$$
where $n$ is the largest integer for which elements 
$a_0 \, a_1 \, \cdots a_n$ produce the same poset (with the same
labeling) as elements $b_0 \, b_1 \, \cdots b_n$.  It is easy to verify
that this yields a metric on $\Omega$; indeed $d$ satisfies a condition
stronger than the triangle inequality: for any three causets ${a}$,
${b}$ and ${c}$, we have $d({a},{c})=\max(d({a},{b}),d({b},{c}))$.  This
``ultrametric'' property derives from the tree structure of the space $\Z$
of cylinder sets (or equivalently truncated histories), as described
earlier.
The maximum distance between two causets is $1/2$ and occurs when their
initial two elements already form distinct partial orders.  Notice also
that the open balls in this metric are exactly the cylinder sets, with
the radius of the ball serving as a measure of ``fineness''.
 
One can see without too much difficulty that with this metric, 
$\Omega$ becomes a compact topological space.\footnote{$^\flat$}%
{Proven explicitly in the next sub-section.}
Moreover the cylinder sets,
being the balls of some radius, 
are both open and closed: clopen.  It
follows by definition that $\Omega$ has a basis of clopen sets and that
every open set is a countable union of cylinder sets (there being only a
countable number of cylinder sets because the finite causets are only
countable in number).  Since each element of the event algebra $\S$ is
itself a (finite) union of cylinder sets, we can conclude that the open
sets are precisely the members of $\bigvee\S$, the closed sets, their
complements, being then the members of $\bigwedge\S$.  The events that
belong to both these families are the clopen events, and they clearly
include all of $\S$, because a finite union of open [respectively
closed] sets is also open [closed].  Let us prove the converse, that
every clopen event belongs to $\S$.

\LEMMA(4.1) \quad $\bigvee\S \; \cap \; \bigwedge \S \; = \; \S$

\PROOF  We are asked to prove that $\S$ comprises precisely the 
clopen subsets of $\Omega$.  Since we already know that every $A\in\S$
is clopen, it suffices to verify that any clopen $A$, i.e. any
$A \in \bigvee\S \, \cap \, \bigwedge \S$~, also belongs to $\S$.  Let
$A \in \bigvee\S$.  By definition $A$ is a union of cylinder sets:
$A=Z_1 \cup Z_2 \cup Z_3 \dots$.  Now this sequence either terminates at
a finite stage or it does not.  If it does terminate then $A$ is a
finite union of cylinder sets, whence a member of $\S$, and we are done.
If it does not terminate, then we can find a sequence of points
$x_j{\in}A$ which escapes from every $Z_j$; and because $\Omega$ is
compact, we can suppose that this sequence converges to some $x\in\Omega$.
This $x$ cannot lie in any given $Z_k$ because it is a limit of points
$x_j$ which eventually belong to the closed set $\Omega\less Z_k$;
consequently $x\notin A = \cup_k Z_k$.  We have thus constructed a
sequence of points of $A$ which converge to a point outside $A$, meaning
that $A$ is not closed.  It thus cannot be clopen.  Or if it is clopen,
then we are back to the terminating sequence and the conclusion that
$A\in\S$.  \qed

Turning to the 2-site hopper, we need to make only one change to what
has been written above for causets.  The histories are now
sequences of digits, 0 or 1, beginning with $0$, and the integer $n$
that occurs in the 
definition (5) is now the largest index such that the
two subsequences 
$(0 \; a_1 \; a_2 \, \cdots \, a_n)$ and
$(0 \; b_1 \; b_2 \, \cdots \, b_n)$ 
coincide.
The rest is all the same.  The history space $\Omega$ is still a compact
metric space, the cylinder sets are clopen and generate the topology,
etc.

\subsection {the tree of truncated histories}
We have already seen in section 2, that a point of $\Omega$, that is to
say a history, can be construed as a path $\gamma$ through the tree $\T$, each
node of which is a ``truncated history'', meaning, as the case may be,
either a finite causet or a finite sequence of binary digits.\footnote{$^\star$}
{The paths under consideration in what follows will usually begin at the
 ``root'' of $\T$ (corresponding to the cylinder-set $\Oga$), but
 sometimes they will originate at some other node of $\T$.  The two
 cases are actually interchangeable because any path not originating at
 the root has a unique extension back to it, $\T$ being a tree.
}
Flowing
from this correspondence between histories and paths through $\T$ is a
different way to characterize certain types of events, including the
open sets $\bigvee\S$ and more generally the events in $\R\bigvee\S$.  

Consider first a cylinder set $Z=\cyl(h)$, where $h\in\Omega(n)$ is a
truncated history.  Which paths $\gamma$ correspond to this
cylinder set?  By definition they are just the paths whose corresponding
histories reproduce $h$ when truncated at the $n^{th}$ stage, that is,
they are precisely the paths that {\it pass through the node} in $\T$
that represents $h$, which I will denote either as $h$ itself or as
$\node(h)$ in order to emphasize that $h$ is being treated as a node in
$\T$.  Because $\T$ is a tree, such a path necessarily follows one of
the branches emanating from $\node(h)$; it then remains forever in the
``upward subset'' of $\T$ consisting of all descendants (in $\T$) of
$h$.  In this way, every open event $A\subseteq\Omega$ can be
represented by an upward-closed subset $\alfa\subseteq\T$, and vice
versa, given such a subset the paths that enter (and consequently remain
in) $\alfa$ comprise an open event $A\subseteq\Omega$.  More generally,
every subset $\alfa\subseteq\T$ gives rise to an event $S(\alfa)$ by the
same rule:

\DEFINITION  $S(\alfa) = \SetOf{\gamma}{\gamma \hbox{ is eventually in } \alfa}$

\noindent
Here, $\gamma$ is a point of $\Omega$, represented as 
a path $\gamma = (h_0, h_1, h_2 \cdots)$ through $\T$,
and the statement that this path is {\it eventually} in $\alfa$ means
that $(\exists n_0)(\forall n>n_0)(h_n\in\alfa)$.
Evidently, $S(\cdot)$ commutes with the Boolean operations: 
$S(\alfa \beta)=S(\alfa)S(\beta)$, 
$S(\alfa \less \beta)=S(\alfa)\less S(\beta)$, 
$S(\alfa + \beta)=S(\alfa)+S(\beta)$, etc.
(where $\alfa + \beta := (\alfa\cup\beta)\less(\alfa\beta)$ is the
Boolean operation of ``addition modulo 2'').

As a further aid to intuition, one can conceive of certain types of
events in terms of ``properties'' acquired or lost in the course of the
process under consideration.  Formally, this corresponds closely to the
characterization by sets of nodes in $\T$, but it carries perhaps a more
``evolutionary'' feeling.
For example consider the event of return analyzed earlier. 
One can cook up a ``property'' which the particle possesses when, and only
when, it has returned to the origin.  
By definition, this property of ``having returned'' is {\it hereditary}
in the sense that, once acquired, it can never be lost.
Topologically,
the set of all paths $\gamma$ which acquire a hereditary property yields
an open subset of $\Omega$, as is easily corroborated if one thinks
through the definitions.
Dually, a property 
that once lost can never be regained, but that every
path begins with, corresponds to a closed set
(causet example: being an originary).
And a property that can
be acquired but never regained if lost yields an event of the form
$A\less{}B$, where both $A$ and $B$ are open.  
(Hopper example: visiting $x=1$ but not $x=2$~. 
 Notice that this third type of property includes both of the
 previous two as special cases.)
In terms of sets of nodes like the sets $\alfa$ discussed above, the
first type of property is an upward-closed subset of $\T$, the second is
a downward-closed subset, and the third is a {\it convex} subset,
defined as a subset of $\T$ that contains, together with nodes $h_1$
and $h_2$, every node that lies on some path from $h_1$ to $h_2$.
In order-theoretic language for the poset $\T$, this
just says that $\alfa$ includes the {\it order-interval}
between any two of its elements.\footnote{$^\dagger$}
{The order-interval delimited by elements $x$ and $y$ of some poset 
 is $\SetOf{z}{x\prec z\prec y}$~.}
In Sections 5 and 6, the events of the form $S(\alfa)$ for some
convex $\alfa$ will be among those for which we will able to produce a
canonical representation as a limit of clopen events.

As an application of some of these ideas, let us prove the assertion
made earlier that $\Omega$ is topologically compact.  
By a standard criterion for compactness, 
it suffices to prove that any covering
of $\Oga$ by cylinder sets has a finite sub-covering, so consider an
arbitrary collection of cylinder sets $Z\in\Z$ that covers $\Oga$.  In
relation to $\T$, such a covering is a collection of nodes which
no path $\gamma$ can avoid forever.  Now the (incomplete) paths that do
avoid these nodes fill out a subtree\footnote{$^\flat$}
{a downward-closed subset of $\T$.} 
~$\T'$ of $\T$~,  with the property that no path $\gamma$ can remain
within $\T'$ forever.  But it is well-known that such a tree can have only
a finite number of nodes, assuming that no node has an infinite
branching number.  (This has been called the ``infinity lemma'' of graph
theory.)  The maximal elements of $\T'$ thus furnish a finite collection
of nodes that every path must encounter.  In their guise as
cylinder-sets these nodes constitute then a finite subcover of $\Oga$. 


\section{5.~Set-theoretic limits of events}                     
The most 
elementary
kind of limit that one can imagine for a sequence
of events partakes of neither metric nor topology nor measure; it is
purely set-theoretic.  We have already seen how increasing sequences of
clopen sets yield the open sets $\bigvee\S$, while decreasing sequences
of clopen sets yield the closed sets $\bigwedge\S$.  In both cases the
operative concept of limit emerges more or less automatically.  Going
beyond these two types of approximation, we can recognize a more general concept
of which $\bigvee$ and $\bigwedge$ are special cases.  Let $X_j$ be a
sequence of subsets of $\Oga$, and deem it to be convergent when we
have for any point $x$ of $\Oga$ that eventually $x\in X_j$ or
eventually $x\notin X_j$.  In such a case, we will write ~$X=\lim X_j$~,
where of course $X$ consists of those $x$ that realize the first
alternative of being eventually in $X_j$~.  
The set of all events obtainable in this
way as limits of events $A_j\in\S$, I will denote as $\Lim\S$~. 
Notice that `$\lim$' commutes with the Boolean operations:
$$
  \lim(A_n \cup B_n) = (\lim A_n) \cup (\lim B_n) \ , \ \hbox{ etc.}
  \eqno(6)
$$

In trying to extend our vector-measure $\mu$ beyond the clopen events,
one might hope that one could at least get as far as $\Lim\S$.  
Were $\mu$ an ordinary measure, this would be true, 
because ~$\lim A_j$~
would be sandwiched between the measurable sets
~$\limsup A_j = \cap_j \cup_{k>j} A_k$~
and
~$\liminf A_j = \cup_j \cap_{k>j} A_k$~,
both of which are equal to ~$\lim A_j$~
when the latter exists.  This would
ensure that ~$\lim A_j$~ was measurable and that
~$\lim \mu(A_j) = \mu(\lim A_j)$~.
But with quantal measures this argument is not available, and it turns
out that convergence can fail already for certain decreasing sequences
of clopen events whose measures diverge [14] to infinity.
On the other hand, convergence succeeds for many other sequences, and
one might hope that the failures were confined to physically
uninteresting questions.

Of course, the failure of convergence in even some cases is likely to
contaminate other cases, making it dubious that $\mu(A)$ can be defined
without some further limitation on the sequence $A_j$ beyond the mere
requirement that ~$\lim{}A_j=A$~.  In the next two sections we will
investigate some restrictions of this sort.  For now, let's notice that
for open sets $A$ there exists a very naturally defined canonical
sequence of events $A_n\in\S_n$ converging to $A$.  Namely, we can take
for $A_n$ the union of all the cylinder sets from $\Z_n$ that are
contained within $A$.  This yields a ``best approximation to $A$ at
stage $n$'' in the sense that $A_n$ couldn't be enlarged without the
sequence losing its increasing nature.

Dually, one immediately obtains a canonical choice of sequence for any
closed event $B$ (just apply complementation to the sequence of clopen
events approximating $\Oga\less{}B$), but having thus two different
classes of canonical sequences 
introduces an ambiguity for events that are both open and closed.
Fortunately, the ambiguity in this case does no harm because
a clopen event necessarily belongs to $\S$, according to Lemma 4.1.
Both the increasing and decreasing canonical sequences thus terminate at
a finite stage: they differ only transiently.



Leaving aside questions of convergence and uniqueness, one might ask how
many events the above limit process can access, even in the best case.
That is, 
how many of the interesting questions even belong to $\Lim\S$ at all?  
With reference to the causal set case, 
recall first of all that one encounters
all the 
stem-events without ever leaving the open sets $\bigvee\S$.
Remembering also that $\Lim\S$ is closed under the Boolean operations,
we can thus say on the positive side
that every finite logical combination of stem-events
is available within $\Lim\S$
(as also the entire event-algebra $\R\bigvee\S$ of course).
On the negative side however, we can notice that
events like the post-event and (for the particle case) the event of
infinite return fall outside of $\Lim\S$, as a consequence of the
following lemma.

\LEMMA(5.1) \quad  Let $A\subseteq\Omega$.
              If both $A$ and $\Oga\less A$ are dense 
              subsets of $\Omega$ 
              then $A\notin\Lim\S$.  

\PROOF   In the following $A\orthog B$ will mean that $A$ and $B$ are disjoint. 
Suppose, for contradiction, that $A=\lim A_n$ with $A_n\in\S$,
and write $\Abar_n$ for its complement $\Oga\less{}A_n$, also taking
note of the fact that $\Abar_n$, like $A_n$ itself, is clopen.
We will find inductively a subsequence $A_{n_1},A_{n_2},A_{n_3},\cdots$
of the $A_n$ and a matched sequence of clopen sets
$B_1\supseteq{}B_2\supseteq{}B_3\cdots$ such that $B_j$ is alternately
included in and disjoint from $A_{n_j}$.

\noindent step 1.  To start with, put $n_1=1$ and $B_1=A_{n_1}$.  We have
$B_1\subseteq A_1$.

\noindent step 2. Next observe that since $B_1$ is open and $\Oga\less A$ is dense, 
there exists $x\in B_1\cap(\Oga\less A)$.  Then since $x\notin A=\lim_n A_n$,
there exists by hypothesis some $n_2>n_1$ such that $x\notin A_{n_2}$,
i.e. $x\in\Abar_{n_2}$.  
Put $B_2=\Abar_{n_2}\cap B_1$, which is again
clopen since both $\Abar_{n_2}$ and $B_1$ are clopen.  
We have $B_2\part B_1$ with $B_2 \orthog  A_{n_2}$.

\noindent step 3.  For step 3, we proceed exactly as in step 2 with the roles of
$A$ and $\Oga\less A$ interchanged.  Namely we observe that since $B_2$
is open and $A$ is dense,  
there exists $x\in B_2\cap A$.  Then since $x\in A=\lim_n A_n$,
there exists by hypothesis some $n_3>n_2$ such that $x\in A_{n_3}$.
Put $B_3=A_{n_3}\cap B_2$, which is again
clopen since both $A_{n_3}$ and $B_2$ are clopen.  
We have $B_3\part B_2\part B_1$ with $B_3\part A_{n_3}$.

\noindent Now proceed inductively to produce $B_4\part A_{n_4}$,
$B_5 \orthog A_{n_5}$, etc.  
Finally put $B=\lim_n B_n= \bigcap\limits_{n=1}^\infty B_n$ and note that $B$ is non-empty since the
$B_n$ are all compact.  (Indeed, every event in $\S$ is compact, being a
closed subset of the compact space $\Oga$.)
Pick any $x\in B$.  
For odd  $j$ we have $x\in B_j\part   A_{n_j}$ $\implies$ $x   \in A_{n_j}$.
For even $j$ we have $x\in B_j\orthog A_{n_j}$ $\implies$ $x\notin A_{n_j}$.
Thus the $A_n$ vacillate between including and excluding $x$,
contradicting our assumption that  $\lim A_n$ exists. \qed

\noindent
The lemma applies to the post-event because, no matter how far the
growth process has proceeded, the growing causet ``still has a free
choice'' whether to end up with a post or without one (and exactly the
same thing can be said for the event of infinite return).  But this
freedom means precisely that both the post-event and
its complement are dense in $\Oga$.

Lemma 5.1 shows that $\Lim\S$ is far from containing every event of
potential interest, but one might wonder exactly how far.  One answer
comes from Exercise (22.17) of reference [12], according to
which $\Lim\S$ equals what is called $\Delta^0_2$, defined to be the
intersection of $\bigvee\bigwedge\S$ and $\bigwedge\bigvee\S$.  This
places $\Lim\S$ at a very low level of the so called ``Borel
hierarchy'', which continues on for $\aleph_1$ steps beyond $\Delta^0_2$
before it exhausts the Borel subsets of $\Oga$.  In this sense the
limiting process `$\lim$' does not take us very far beyond the clopen
events.
On the other hand, we have also seen that by applying `$\lim$' more than
once, one can reach, for example, the post-event.  How many events can
one reach in this manner?  When combined with other results in
[12], (22.17) therein also answers this question by implying
that (transfinite but still countable) iteration of the `$\lim$'
operation suffices to produce any Borel set.  In this sense the $\lim$
operation is quite far reaching, given that an event of interest but not
falling within the Borel domain would be hard to conceive of.

\section{6.~Canonical approximations for certain events}	
For an event $A\part\Oga$ which is open with respect to the topology
defined in Section 4, that is for $A\in\bigvee\S$, we have already
discovered one canonical sequence $A_n$ of approximations to
$A$.  
The cylinder sets $\Z_n$ ``at stage $n$'' provide a kind of ``mesh'' in
$\Oga$ whose fineness increases with $n$, 
and our canonical choice of approximating event at stage $n$ was
$$
  A_n=\bigcup\SetOf{Z\in\Z_n}{Z\part A}  \ , \eqno(7)
$$
the biggest member of $\S_n=\R\Z_n$ which can fit inside $A$.
As we have seen,
the $A_n$ converge to $A$ in the sense defined in section 5,
but of course there exist many other sequences $B_n\in\S_n$ which also
converge to $A$ in this sense, and when the vector-measure $\mu$ is
not of bounded variation, there is no guarantee that the corresponding
sequences $\mu(A_n)$ and $\mu(B_n)$, if they converge at all, will
converge to the same limit.  In general they doubtless will not if $B_n$
is chosen with sufficient malice.
In the face of such ambiguity, one might still hope to find some
reasonably inclusive event-algebra $\A\supseteq\S$ and for each event $A\in\A$ a
canonical approximating sequence of events $A_n\in\S_n$ with $\lim A_n = A$ and
such that $\mu(A_n)$ was a convergent sequence in Hilbert space.  The vector
$\lim_n\mu(A_n)$ could then be adopted as the definition of $\mu(A)$.

One snag that this perspective encounters is apparent already for the
case where we are approximating open sets $A$ and $B$, and our canonical
approximations $A_n$ and $B_n$ are the ones 
given by (7).
From
$\lim A_n = A$ and 
$\lim B_n = B$ 
it does indeed follow, 
as we have already noted, 
that 
~$\lim (A_n \cap B_n) = A \cap B$~ and
~$\lim (A_n \cup B_n) = A \cup B$~.
For the case of intersection it even follows
that the events ~$(A_n \cap B_n)$~ provide the canonical approximations to
the open event $A\cap B$, but
the analogous conclusion fails for the case of union because the
canonical approximation $(A\cup B)_n$ will in general be larger than
$(A_n \cup B_n)$, since some cylinder set $Z\in\Z_n$ can, by
``straddling the boundary'' between $A$ and $B$, be included in
$A\cup{B}$ without being included in either $A$ or $B$.  One would thus
obtain different approximating sequences for $A\cup{}B$, depending on
whether one regarded it as an open set in its own right or as the
result of uniting $A$ with $B$.  In the next section we will begin
to see what it would take to render this kind of ambiguity harmless.
For now however, let's ignore that issue and consider simply the
question of finding unambiguous approximating sequences for as many
members of $\R\bigvee\S$ (= $\R\bigwedge\S$) as possible.

To that end, let's return to the tree $\T$ of truncated histories and
the method of representing certain events by subsets $\alfa\part\T$.
Although we didn't make it explicit earlier, it is clear that 
a sequence of events $A_n\in\S_n$
is equivalent to 
a set of nodes $\alfa\part\T$. 
Indeed, each $A_n$ is a union of cylinder sets $Z\in\Z_n$, and
each such cylinder set corresponds to a node in $\T_n$, the $n^{th}$
level of $\T$.
This associates to each $A_n$ a set of nodes at level $n$, and
amalgamating the nodes of all levels into a single collection yields
$\alfa$.
Conversely, given $\alfa\part\T$ we obtain $A_n$ as the union of the
cylinder sets that correspond to the $\alfa$-nodes at level $n$.
Since the correspondences between 
cylinder sets $Z\in\Z_n$, 
nodes in $\T_n$, and 
truncated histories $\gamma\in\Oga(n)$ 
are so close, 
I will often identify all three with one another, speaking for
example of a cylinder set $Z$ as a node in $\T$.   
When this is done, we can express the correspondence between
node-sets $\alfa$ and approximating sequences $(A_n)$ 
in a simple formula
by writing
$A_n=\bigcup(\alfa\cap\Z_n)$.

Now let $\alfa$ be any set of nodes and let the $A_n$ be the corresponding
sequence of events.  Recall that we defined 
$S(\alfa)$ as the event that $\gamma$ is {\it eventually} in $\alfa$:
$$S(\alfa)=\SetOf{\gamma}{(\exists N)(\forall n>N)(\gamma_n\in\alfa)}\ .$$
Dually one can also define $\Stilde(\alfa)$ as the event that 
$\gamma$ is {\it repeatedly} in $\alfa$:
$$\Stilde(\alfa)=\SetOf{\gamma}{(\forall N)(\exists n>N)(\gamma_n\in\alfa)} \ .$$
It follows, simply by tracing through the definitions, that 
$$
\eqalign{
        S(\alfa)=\liminf A_n  \ ,    \cr
  \Stilde(\alfa)=\limsup A_n  \ .  
}
\eqno(8)
$$
Since ~$\lim A_n$~ exists if and only if ~$\liminf A_n=\limsup A_n$~ (in
which case their common value equals ~$\lim A_n$), we learn that the
events of the form ~$\lim A_n$~  are precisely those for which 
~$S(\alfa)=\Stilde(\alfa)$~, 
which in turn are precisely those such that 
no path $\gamma$ 
can leave and re-enter $\alfa$ more than a finite number of times.  
Evidently this property generalizes the concept of convexity which we
met with earlier.
Notice incidentally that 
equations (8)
imply that the forms
$S(\alfa)$ and $\Stilde(\alfa)$ don't reach beyond $\bigvee\bigwedge\S$
and $\bigwedge\bigvee\S$, known in descriptive set theory as
$\Sigma^0_2$ and $\Pi^0_2$,
respectively.  
Roughly, they reach as far as events whose complexity is that of
the post event.
Very optimistically, one might hope to 
go beyond this and
find for any Borel set
$A\part\Oga$, some sort of canonical presentation in terms of clopen
events,
but in this section we will not venture outside of
$\R\bigvee\S$, the finite Boolean combinations of opens.
Since $\R\bigvee\S\part\Lim\S$, all such events can be expressed as
$S(\alfa)$ for some subset $\alfa\part\T$.

What we are asking for is a sort of ``normal form'' for events $E$
in $\R\bigvee\S$.  As a first step in that direction, let us prove that
every such event can be expressed as a disjoint union of events, each of
which has the form, 
open$\less$open,
or equivalently, 
open $\cap$ closed.

\LEMMA(6.1) \quad  Let $E\in\R\bigvee\S$ be a finite logical combination of open events.  
  Then there exists a decreasing sequence of open events
  $E^1 \supseteq E^2 \supseteq E^3 \cdots \supseteq E^{K}$ such that 
  $E = E^1 + E^2 + E^3  \cdots + E^{K}$
  $ = E^1\less E^2 \, \sqcup \,  E^3\less E^4 \, \sqcup \cdots$, 
  where `$\sqcup$' denotes disjoint union.
  Moreover the $E^j$ are formed from the original events using
  only the operations of union and intersection.

\PROOF  In this proof, as in the statement of the lemma, we use the
   operation of Boolean addition, 
$$
   A+B = (A\cup B)\less (A\cap B) \ ,  \eqno(9)
$$ 
and we write the intersection of two sets as their product.  
   Any Boolean combination of sets is
   then a polynomial in these sets, and since products of open sets
   are open, any Boolean combination of open events can be expressed
   simply as a Boolean sum of open events.  
Given these facts, a 
   proof by induction is not hard to devise, but 
   it seems clearer just to
   illustrate the pattern involved with the cases of $K=2,3$.  
   For two events we have 
   $A+B=(A+B+AB)+AB=A\cup B + AB$.  
   For three we have
   $A+B+C
   =A+(B+C)
   =A + (B\cup C + BC)
   =(A + B\cup C) + BC
   =(A \cup B\cup C + A (B\cup C)) + BC
   = A\cup B\cup C + ( A(B\cup C) + BC )
   = A\cup B\cup C + A(B\cup C)\cup BC + A(B\cup C)BC
   = A\cup B\cup C + (AB\cup AC\cup BC) + ABC$.
   The ``inclusion-exclusion'' pattern that is evident here emerges
   with particular clarity
   when one interprets Boolean addition as addition of
   characteristic functions modulo 2. 
   The final equation in the statement of the
   lemma then follows directly from the fact that the $E^j$ are
   decreasing. 
   One can also restate the essence of the proof in a simple formula:
   $\sum_{j=1}^K A_j = \sum_{j=1}^K B_j$, where 
   $B_j=\SetOf{x}{x \hbox{ belongs to at least $j$ of the } A_k}$,
   this being clearly a union of intersections of the $A_k$.
  \qed
    
Given any set $E$ expressed as in the lemma, we get immediately the
approximations $E_n = E^1_n + E^2_n + \cdots + E^K_n$, 
where $E^j_n$ is our canonical $n^{th}$
approximation to the open set $E^j$, 
and thence the corresponding sets of nodes 
$\alfa_n=\alfa^1_n+\alfa^2_n+\cdots+\alfa^K_n$ together with their union
$\alfa=\cup_n\alfa_n$. 
However, this construction is only a first step toward uniqueness,
because the resulting $\alfa$ still depends on the original choice of
the $E^j$, which are not given to us uniquely by the lemma.

In working toward a unique approximating sequence, let us 
concentrate on
the simplest case
of an event $E = A \less B$ which is the difference of only two open
sets $B\part A$ (corresponding to $K=2$ in the lemma).
Can we render $A$ and $B$ unique in this case?
It's not difficult to demonstrate that if we gather together all pairs
$A\supseteq B$ such that $E=A\less B$, then the union of all
the sets $A$ and the union of all the sets $B$ yields another such pair.
Evidently this ``biggest pair''  is unique and uniquely
determined by the original event $E$.  This in turn yields 
[by (6)]
a canonical
sequence of approximations $E_n$ to $E$ of the form, $E_n=A_n\less B_n$,
where $A_n\in\S_n$ and $B_n\in\S_n$ are 
the canonical $n^{th}$
approximations to $A$ and $B$.  In terms of the equivalent node-sets $\alfa_n$ 
these approximations are given by $\alfa_n\less\beta_n$, whose union
over $n$
I'll designate simply by $\alfa$, following our earlier notation.

Although the node-set $\alfa$ that we have found 
is canonical and concisely defined,
one might wish for a more constructive route to it, or at least a
characterization of it in terms of more easily verifiable necessary and
sufficient conditions.  The remainder of the present section will
develop a prescription of this sort.  In fact, I am not certain that the
second prescription will be strictly equivalent to the first.  
If it is, that is all to the good since we will then not be forced to
choose between the two.  
If it isn't, that
doesn't really matter, since the second prescription stands on its own
and, being  more concrete, is likely to be more useful in practice.

\LEMMA(6.2)~ If $\alfa$ and $\beta$ are upward-closed subsets of $\T$ with
$\alfa\supseteq\beta$ then $\alfa\less\beta$ is convex.

\PROOF  We are to show that no path between two nodes $x$ and $y$ in
$\alfa\less\beta$ can contain nodes outside of $\alfa\less\beta$.
Equivalently, no path which has left $\alfa\less\beta$ can ever re-enter
it. 
But since $\alfa$ is upward-closed 
no path from $x\in\alfa$ can leave $\alfa$, 
therefore it can leave $\alfa\less\beta$ only by entering $\beta$; 
it then must remain in $\beta$ (which is also upward-closed)
forever,
and consequently can never re-enter $\alfa\less\beta$.  \qed

Now let $E=A\less B$ as above and let $\alfa$ and $\beta$ be the
corresponding node-sets.  Since $A$ and $B$ are open, both 
$\alfa$ and $\beta$ are upward-closed subsets of $\T$.  We also know
that $A=S(\alfa)$, $B=S(\beta)$ and $A\less B=S(\alfa\less\beta)$.  The lemma
then teaches us that $A\less B=S(\hat\alfa)$, with $\hat\alfa$ a convex subset
of $\T$.  
The converse is true as well:

\LEMMA(6.3)~  If $\hat\alfa\part\T$ is convex then $S(\hat\alfa)=A\less B$ for some open
        sets $A$ and $B$.

\PROOF  Recalling that we have identified points of $\Omega$ with
infinite paths $\gamma$ through $\T$, let $A$ be the set of all paths
that enter $\hat\alfa$, and let $B$ be the subset of these that
subsequently leave $\hat\alfa$.  
By definition $S(\hat\alfa)=A\less{}B$,
but both $A$ and $B$ are open because  
the property of ``having entered $\hat\alfa$'' and
the property of ``having left $\hat\alfa$'' are both hereditary. \qed

\noindent
Henceforth, we will just deal with the convex subset $\hat\alfa$,
renaming it to plain $\alfa$ for simplicity.  That is, we will be
concerned with a fixed event $E$ of the form  (open$\less$open) and with
a convex set of nodes $\alfa\part\T$ such that\footnote{$^\star$}
{In view of (8) we would also want in general to require that
$S(\alfa)=\Stilde(\alfa)$, but this holds automatically when $\alfa$ is
convex.}
 $E=S(\alfa)$. 

Let us say that $\alfa\part\T$ is {\it{}prolific} if it lacks maximal
elements. 
A second requirement that adds itself very naturally to convexity is the
condition that $\alfa$ be prolific in this sense.  Given convexity, this
is equivalent to saying that every node $x\in\alfa$ originates a path
that remains forever within $\alfa$.
In the opposite case, $\alfa$ will contain ``sterile'' nodes from which
all paths eventually leave $\alfa$ for good.
It is clear that removing these sterile nodes will not alter $E$, nor
will it spoil the convexity of $\alfa$.  
We can therefore always arrange that $\alfa$ be both convex and
prolific.
The ``pruning'' of the ``sterile'' nodes in order to render $\alfa$
prolific also appears as a very natural operation when it is expressed
in terms of cylinder sets $Z$.  It simply removes from $\alfa$ those $Z$
which are disjoint from $E$.
%

We have now arranged for $\alfa$ to be convex and prolific, but this
does not yet make it unique, since for example we could remove all the
nodes up to any fixed finite level $n$ without altering $S(\alfa)$.  
If we did so, however, we might create a situation where, for example,
some cylinder set $Z$ was wholly included in $E$ without $Z$ itself
(regarded as a node in $\T$) belonging to $\alfa$.
To remedy this kind of lacuna, we can adjoin to $\alfa$ 
every node $Z$ such that every path originating at $Z$
eventually enters $\alfa$.  
It is again easy to see that adjoining these nodes will not interfere
with $\alfa$ being convex and prolific.

In this last step we have, in a manner of speaking, completed $\alfa$
toward the past, but in fact there is cause to carry this process of
``past-completion'' farther by adjoining still other nodes to $\alfa$.
These additional nodes are perhaps not such obvious candidates as 
the previous ones,
but throwing them in as well 
(which I think corresponds to enlarging the open set $A$) 
will provide us with the uniqueness we are seeking.

\DEFINITION Let $x\in\T$ and $\alfa\part\T$.  Then $x\prec\alfa$ means that
$x$ precedes some node in $\alfa$: $(\exists y\in\alfa)(x\prec y)$~.

\REMARK In terms of cylinder sets, \ 
 $Z_1\prec Z_2 \iff Z_2\supseteq Z_1$~.

\DEFINITION  The {\it exclusive past} of $\alfa$ is the set of nodes
      strictly below $\alfa$: $\SetOf{x\notin\alfa}{x\prec\alfa}$~.

\noindent
Using this definition, let us say that $\alfa$ is {\it past-complete}
if its exclusive past $P$ is prolific, which in turn says that any
node in $P$ originates a path that repeatedly visits $P$.
I claim we can render $\alfa$ past-complete be adjoining to it all nodes
below $\alfa$ that fail to satisfy this last condition, and furthermore
that the resulting set of nodes $\alfa'$ 
will yield the same event $E$ as $\alfa$ and 
will be convex and prolific if
$\alfa$ itself was.

\LEMMA(6.4)~  Let $\alfa\part\T$ be any set of nodes and let $\alfa'$ be its
  ``past-completion'' as just described.  Then $\alfa'$ is past-complete.
   Moreover $S(\alfa')=S(\alfa)$ and $\Stilde(\alfa')=\Stilde(\alfa)$~.

\PROOF  That $S(\alfa')\supseteq S(\alfa)$ is obvious.  
   To prove that they are equal it suffices to show that no path can
   eventually remain within $\alfa'$ without also remaining eventually
   within $\alfa$.  Suppose the contrary, and let $\gamma\in S(\alfa')$
   be a path which is repeatedly outside $\alfa$.  By passing to a tail
   of $\gamma$ we can suppose that it is always within $\alfa'$.  Let
   $x\in\gamma$ be a node which is not in $\alfa$ and let $y\in\gamma$
   be a later node of the same type.  Since $y\in\alfa'\less\alfa$ it is
   by definition in $P$, the exclusive past of $\alfa$.  Hence $x$
   originates a path (namely $\gamma$) which visits $P$ at $y$; and
   since there are an infinite number of nodes like $y$, $\gamma$ visits
   $P$ repeatedly.  But this contradicts the criterion for having
   included $x$ in $\alfa'$ in the first place.

\noindent
   The proof that $\Stilde(\alfa')=\Stilde(\alfa)$ is similar.  It
   suffices to show that every path in $\Stilde(\alfa')$ visits $\alfa$
   repeatedly.  Suppose the contrary, and let $\gamma\in\Stilde(\alfa')$
   be a path which is eventually outside $\alfa$.  By passing to a tail
   we can suppose that $\gamma$ is always outside of $\alfa$.  Let
   $x\in\gamma$ be a node which is in $\alfa'$ and let $y_1,
   y_2, \cdots$ be a sequence of later nodes of $\gamma$ which are also
   in $\alfa'$.  Since the $y_j$ belong to $\alfa'\less\alfa$ they are
   by definition in $P$, the exclusive past of $\alfa$.  Hence $x$
   originates a path which returns repeatedly to $P$, contradicting the
   criterion for having included $x$ in $\alfa'$ in the first place.

\noindent
   To complete the proof, we need to show that $\alfa'$ is
   past-complete.  To that end let $x$ be in the exclusive past of
   $\alfa'$.  Since any $y$ in $\alfa'$ is either in $\alfa$ or in its
   exclusive past, and since $\alfa'\supseteq\alfa$, $x$ is also in the
   exclusive past of $\alfa$.  Consequently, since $x$ was not put into
   $\alfa'$, it originates a path $\gamma$ that repeatedly visits the
   exclusive past of $\alfa$.  But by definition, no node in such a path
   would have been put into $\alfa'$ either, whence $\gamma$ repeatedly
   visits the exclusive past of $\alfa'$, as was to be shown.  \qed
 
We also wish to prove that past-completion preserves the attributes of
being prolific and convex.  The first is easy because past-completion
only adds nodes which are below some element of $\alfa$, and this can
introduce no new maximal element.  
  
For the second, we need to demonstrate\footnote{$^\dagger$}
{The demonstration that follows seems somehow longer than it ought to
 be.  Intuitively it suffices to observe first that $\alfa'$ is built up from
 $\alfa$ by successive adjunction of maximal elements of its exclusive
 past, and second that adjoining such an element cannot spoil convexity.}
that if $\alfa$ is convex, and if
$x\prec y$ are nodes in $\alfa'$ then the order-interval $I$ delimited
by $x$ and $y$ is also within $\alfa'$.  
When $x\in\alfa$ the proof is simple, since $y$ is either within $\alfa$
itself or precedes some element $z$ which is.  In either case the $I$ is
included in some second interval $I'$ (possibly the same as $I$) with
endpoints in $\alfa$.  Then $I'\part\alfa$ because $\alfa$ is convex,
whence also $I\part{I'}\part\alfa\part\alfa'$, as desired.
The remaining possibility is that $x\in\alfa'\less\alfa$, in which case
it seems more convenient to deal with paths rather than intervals.  From
the definition of convexity, proving that $\alfa'$ is convex amounts to
showing that no path originating from $x$ can leave $\alfa'$ and then
re-enter it.  
First, observe that since every node of $\alfa'$ precedes some node of
$\alfa$, no node of $\alfa'\less\alfa$ can follow a node of $\alfa$; for if
it did, it would also lie within the convex set $\alfa$.  In
consequence, any path that exits $\alfa$ permanently exits $\alfa'$ as
well. 
Now let $\gamma$ be any path originating from $x\in\alfa'\less\alfa$,
and as before write $P$ for the exclusive past of $\alfa$.  By the
definition of $\alfa'$, every path from $x$ must eventually leave $P$.
%
%
If it does so by
leaving the past of $\alfa$, 
$\SetOf{x\in\T}{x\prec\alfa}$, 
then it certainly can never re-enter $\alfa'$.  
If it does so by entering $\alfa$, then it can exit $\alfa'$ only by
exiting $\alfa$, in which case it can never re-enter $\alfa$ or (as we
just observed) $\alfa'$.

So far, we have established the existence, for our event $E$, of a
node-set $\alfa$ which is convex, prolific, and past-complete.  Let us
complete the story by proving that $\alfa$ is also unique.
To that end, let $\alfa$ and $\beta$ be two  prolific, convex and
past-complete node-sets such that $S(\alfa)=S(\beta)$.  Does it follow
that $\alfa=\beta$?
In demonstrating that the answer is ``yes'', 
I will use the ad hoc notation $\P(\alfa)$ for the exclusive past of
$\alfa$ as
defined earlier,  
with $\Pbar(\alfa)=\alfa\cup\P(\alfa)$ = $\SetOf{x\in\T}{(\exists y\in\alfa)(x\preceq y)}$
being the {\it inclusive past}.

First, let us establish that $\alfa$ and $\beta$ have equal inclusive
pasts: $\Pbar\alfa = \Pbar\beta$.
In fact if $x\in\Pbar\alfa$ then $x$ originates a path $\gamma$ that
visits $\alfa$.  Since $\alfa$ is prolific this path can be arranged to
visit $\alfa$ repeatedly, and since $\alfa$ is convex, such a path can
never leave $\alfa$.  Hence $\gamma\in S(\alfa)$, implying in particular
that $\gamma$ visits $\beta$, whence $x\in\Pbar\beta$.  The converse
follows symmetrically.

Now suppose for contradiction that there exists $x\in \alfa\less\beta$~.  
Such an $x$ belongs by definition to $\Pbar\alfa$, hence to
$\Pbar\beta$, hence to $\P\beta$, which in turn is prolific by 
definition of past-completeness.  
Thus $x$ originates a path $\gamma$ that repeatedly visits $\P\beta$.
I claim that $\gamma$ must leave $\alfa$ at some stage.  
(Otherwise 
 $\gamma\in S(\alfa)$ $\implies$ $\gamma\in S(\beta)$ 
 $\implies$ $\gamma$ eventually in $\beta$, 
whence $\gamma$ could never again visit $\P\beta$~.)
And since $\alfa$ is convex, $\gamma$ must remain outside of
$\alfa$ once it has left.  
On the other hand, $\gamma$ must continue to visit $\P\beta$,
which in turn is a subset of $\Pbar\beta=\Pbar\alfa$.
But if $\gamma$ really visited some $y\in\Pbar\alfa$, then by definition 
we could divert it at $y$ to some other $\gamma'$ that would re-enter
$\alfa$, something that we just proved to be impossible.  
This completes the proof of:

\THEOREM(6.5)~  Every event $E$ of the form $E=A\less B$ with $A$ and $B$ open
can be expressed as $E=S(\alfa)=\Stilde(\alfa)$ for a unique set of
nodes $\alfa$ which is convex, prolific and past-complete.
\smallskip

The theorem furnishes a canonical set $\alfa$ of nodes corresponding to
$E$, and as explained earlier, one obtains immediately from such an
$\alfa$ a canonical sequence of approximants $E_n$ to $E$ such that
$E=\lim_n{}A_n$.  We have thus reached our immediate goal.

The canonical approximating sequences of the theorem provide a good
reference point for further developments, and we have learned how to
arrive at them 
step by step, 
starting from the open sets $A$ and $B$.
Nevertheless it seems unlikely that we can limit ourselves to these
particular approximants in general.
Rather, as remarked already at the beginning
of this section, one will in general have to deal with many different
sequences converging to the same event, unless perhaps 
it is possible to
devise
canonical sequences which are closed under the Boolean operations.

We already encountered an ambiguity of this nature when we noticed that
our original,
increasing
canonical approximants (7)
for open events 
(call
them ``C1'') 
are not fully compatible with the Boolean operation of
complementation.  Specifically for a clopen event $E$, these C1
approximants depend on whether one derives them directly from $E$ or by
complementing the corresponding approximants for the open event
$\Oga\less{}E$.
We run into a further, but related conflict if we now compare the C1
approximants with those of the above theorem (call them ``C2'').  
For an
open event $E$ the C1 approximant $E_n$ is nothing but the biggest
member of $\S_n$ included within $E$.  But if we view $E$ as the
difference $E=A\less B$, with $A$ being $E$ itself and $B=0$ being the
empty event, the theorem provides a different set of approximants
$E_n$.  In general, the two disagree, as one can appreciate if one
notices that the pair $(E \; 0)$ is not the ``biggest one'' yielding
$E$.
 
Consider for example the 2-site hopper event that the particle does not
remain forever at its starting site 0, but that the first time it hops
to site 1 it immediately returns to 0.  This event is a union of
cylinder sets corresponding to truncated trajectories of the shape 
$(0, 0, 0, \cdots 0, 1, 0, *)$ 
where the star
`$*$' 
represents any finite sequence of zeros and ones.  
For this event, the node-set $\alfa_1$ of type C1 consists of precisely
the truncated trajectories just indicated.
But that set of nodes is not past-complete.  
Its
completion, the type C2 node-set $\alfa_2$, contains in addition the
truncated trajectories $(0, 0, 0, \cdots 0, 1)$.
Notice that $\alfa_2$ differs from $\alfa_1$ by an infinite number of
nodes in this case.  
(Figure \Fnne {} illustrates this phenomenon.)

\vbox{
  \bigskip
  \includegraphics[scale=0.4]{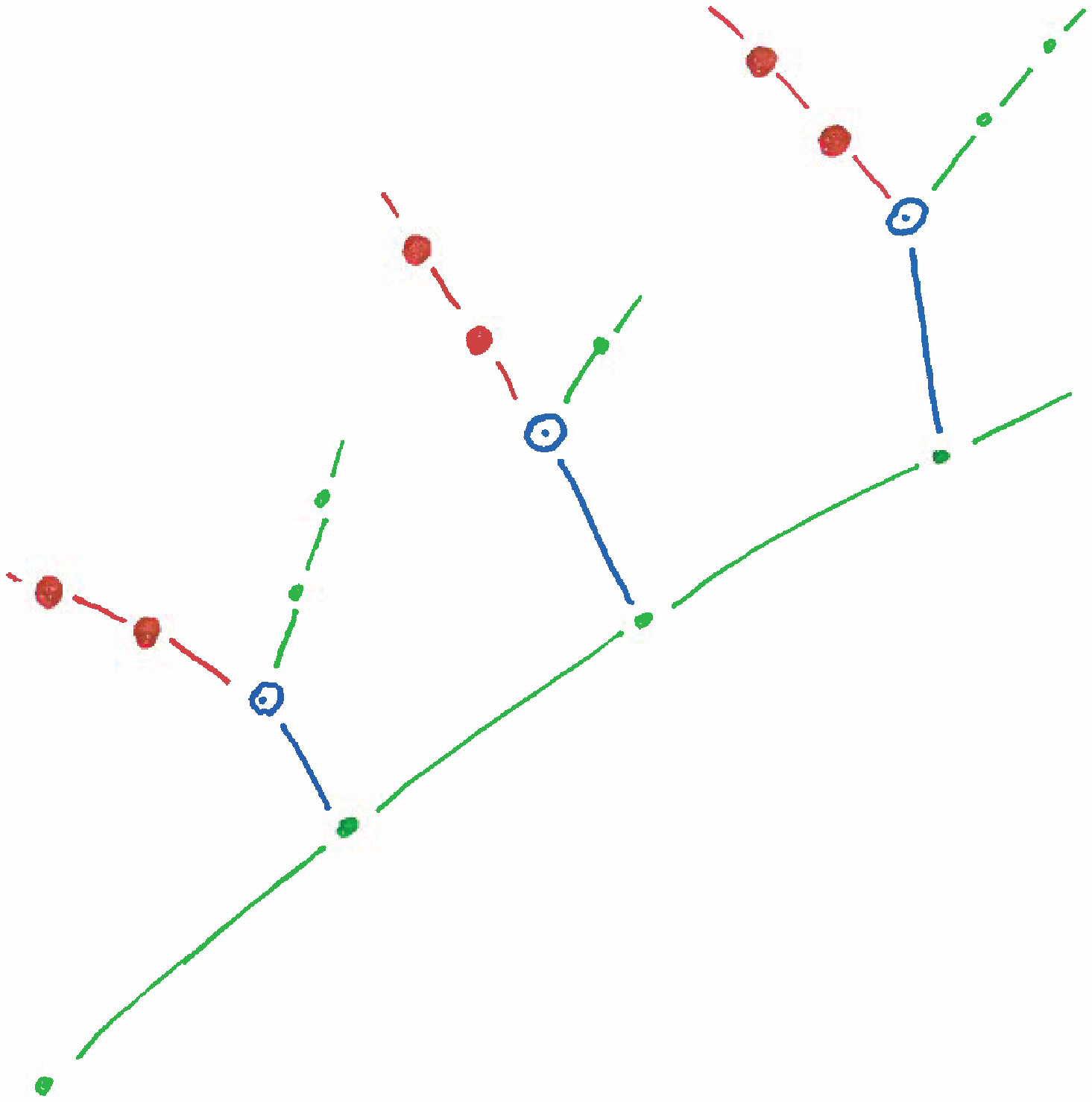}

  \Caption{{\it Figure \Fnne.} 
  Illustrating past-completion in the tree shown.
  The nodes circled in blue ``complete'' those in solid red.  
  Let the node-set be $\alfa_1$  before completion and $\alfa_2$ after completion.
  Evidentally $S(\alfa_1)=S(\alfa_2)$ but only $\alfa_1$ is upward-closed.}}

We thus have
to reckon with
overlapping but in general incompatible prescriptions for
different types of events.  
If one prescription were to be adopted
exclusively,  
it should probably be C2, 
which covers more events than C1 does.
(Incidentally, C2 resolves the aforementioned ambiguity in the C1 prescription in favor
of treating clopen events as closed, not open.)  
On behalf of C1 one might make
the counter-argument that monotonic convergence of the approximants is
to be preferred, but this does not seem so compelling in the context of a
{\it quantal} measure, which itself is not a monotonic set-function.
Better than either choice, however, would be not having to choose at all
because the alternative approximating sequences would all lead to the same
extension of our initial quantal measure.
The main thing for now is that we've discovered at least one canonical
choice of clopen events $E_n$ converging to any event of the form
$E=A\less B$ with $A$ and $B$ open.
     
In the face of these various ambiguities it seems well to emphasize that
none of them affect, in the causet case, the stem-events themselves,
essentially because the latter are not only open but dense in $\Oga$, or
more physically because any growing causet that has not yet produced a
given stem always retains a choice whether or not to do so.  It follows
that not only does the C1 prescription coincide with the C2 prescription
for stem-events (its exclusive pasts being already prolific), but also
the ``biggest pair'' prescription with which we began provably agrees
with the C1 prescription.
The same ought to apply to finite unions and intersections of
stem-events, and similar comments could be made about the event of
``return'' in the hopper case.


Let me conclude this section by sketching very briefly how one might try
to carry our successful ``canonization'' of $E$ = open$\less$open over
to the general case where $E = E^1 + E^2 \cdots + E^K$, the $E^j$ being
open and nested.
Just as earlier we found a ``biggest pair'', $A\supseteq B$, by forming
unions of the individual events $A$ and $B$, one can do the same thing
here with the $E^j$ to obtain (at least in principle) a canonical set
of ``biggest'' 
open and nested 
events $E^j$ such that $E = E^1 + E^2 \cdots + E^K$.
Associating to each such $E^j$ its 
canonical approximants $E^j_n$ (in the C1 sense, say)
then yields for $E$ itself
the approximants $E_n = E^1_n + E^2_n \cdots + E^K_n$~
such that  ~$\lim E_n = E$~.
In principle this achieves our goal, but it remains once again at a
rather abstract level.

As before, we can attempt a more constructive development by working
with the node-set $\alfa\part\T$ corresponding to our approximating
sequence $E_n$, or perhaps with some similar node-set whose uniqueness
can be established directly, and for which we can prove that
$E=S(\alfa)=\Stilde(\alfa)$~.
But how would our construction of $\alfa$ go in this more general case,
and what would generalize the conditions that $\alfa$ be convex,
prolific and past-complete?
It is clear from Lemmas (6.2) and (6.3) that convexity is now too
restrictive.  In its place, one would probably put the more general
requirement that no path $\gamma$ could enter and leave $\alfa$ more
than $K$ times.  Correspondingly one might then expect that $\alfa$
would decompose into subsets $\alfa^j$ that were convex in the strict
sense.
One might also try to arrange for each $\alfa^j$ to be prolific and
past-complete, hoping that this would again confer uniqueness on the
whole collection.
If all this worked out, one would have constructed a canonical
approximating sequence for any Boolean combination of open events, in
particular for any Boolean combination of stem events.


A next step beyond $\R\bigvee\S$, if one could take it, would be to
devise canonical approximations for larger families of events, starting
with the collections ~$\bigvee\bigwedge\S$~ and ~$\bigwedge\bigvee\S$~
in which the post-event and its complement are to be found.  An event
$A$ in either of these collections is accessible from $\S$ as a limit of
limits, but such a double limiting process can only make the potential
ambiguities worse.  
For example, the event $R^\infty$ of repeated return
is in $ \bigwedge\bigvee\S$.  It could be expressed as ~$\lim A_n$~,
where $A_n$ = ``returns at least $n$ times'', or it could be expressed
instead as ~$\lim B_n$~, where $B_n$ = ``returns at least once after
$t=n$''.  Both $A_n$ and $B_n$ give decreasing sequences of open events
and both converge to $R^\infty$, but which sequence, if either, should be
favored as canonical?  Perhaps in certain cases, one could arrive at a
canonical presentation by generalizing further our treatment above in
terms of sets of nodes $\alfa\part\T$, but beyond this, it's not easy to
guess how one might proceed.  To devise canonical approximations for
events of still greater complexity would seem to demand a fresh
approach.


Finally, it might bear repeating here that uniqueness in and of itself
does not guarantee compatibility with the Boolean operations.  And I
believe in fact that none of the prescriptions that this section has
considered are compatible with the full set of such connectives,
albeit some are compatible, for example, with complement or disjoint
union (cf. figure \Ftano).
If there did exist a compatible prescription --- or even a
prescription compatible ``modulo initial transients'', which is just as
good --- that would weigh very heavily in its favor.


\vbox{
  \bigskip
  \includegraphics[scale=0.4]{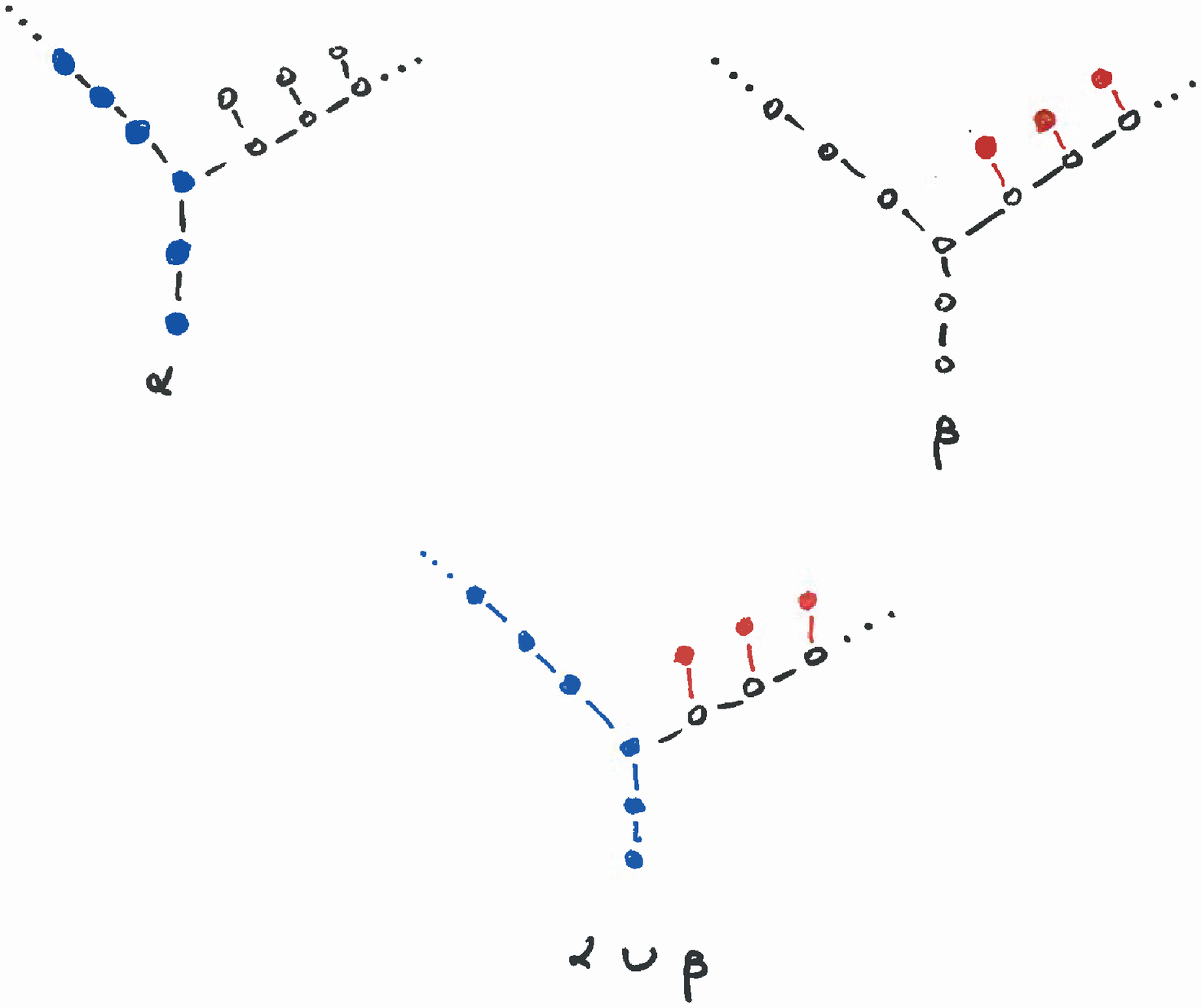}

  \Caption{{\it Figure \Ftano.} 
  Two sets of nodes shown in red and blue.  Both sets are
  convex, prolific and past-complete, but their union is not convex. }}

\section{7.~\Evenlyconvergent sequences of events }
We are given a vector-valued measure $\mu:\S\to\H$ defined initially on
the finite unions of cylinder-events, and we wish to enlarge this
initial domain $\S$ so that it can embrace events like the stem-events
and some of the other events we have been using as illustrations.  In
attempting such an extension or ``prolongation'' of $\mu$, it is natural
to think in terms of approximations, or more formally of limits.
Let $A\part\Oga$ be some event $A$ outside the initial domain.  In order
to define $\ketvector{A}\ideq\mu(A)$~ as a limit, one would aim to identify a
sequence ~$A_n\in\S_n$~ of ``best approximations to $A$'' 
%
%
and one would then hope that the corresponding measures
~$\ketvector{A_n}\in\H$~ would also converge.  If they did, then one
would take their limit in $\H$ to be the measure of $A$:
$$
 \ketvector{A} = \lim\limits_{n\to\infty} \ketvector{A_n} \ .
$$

Notice here that in attempting to define $\ketvector{A}=\mu(A)$ this
way, we have relied on two independent notions of convergence, first the
purely set-theoretic convergence of $A_n$ to $A$ in the sense of section
5 above, and second the topological convergence of the measures
$\ketvector{A_n}$ to $\ketvector{A}$ in Hilbert space (say in the norm
topology or perhaps the weak topology).  One might question whether the
first notion is really needed, given that the extension theorems of
ordinary measure theory do without it, relying solely on the measure
$\mu$ itself.  Would it be possible to proceed similarly here?
Unfortunately, this looks dubious, even though the vector
$\ketvector{A_n}$ carries a certain amount of information about the
event $A_n$ (a very limited amount since, owing to quantal
interference, very different events can share the same vector-measure.)

In the ordinary setting, where $\mu$ is real and positive, it defines a
distance on the space of initially measurable sets modulo sets of
measure zero, such that two events $A,B\in\S$ are close when $\mu(A+B)$
is small.  Extension of the measure then corresponds to completion of
the metric space thereby defined [1].  Quantally,
however, an event of small or zero measure is not negligible in the same
way as it is classically, because of interference.  Thus, if we tried to
use the norm of $\ketvector{A+B}$ as a distance, it wouldn't even obey
the triangle inequality.  (Example: three disjoint events $A$, $B$, $C$
as in the 3-slit experiment of [19] [20];
$\ketvector{A+B}=\ketvector{B+C}=0$ but $\ketvector{A+C}\not=0$.)
Similarly trying to quotient the event-algebra by the events of measure
zero would yield nonsense; it can even happen that all of $\Oga$ is
covered by events of measure zero [17].
To establish an association between a vector $v=\lim\ketvector{A_n}$ and
a definite event $A$ in $\Oga$, one thus seems to need an independent
notion of convergence like that introduced above in section 5 and
developed in section 6.

Accepting this apparent necessity,
let us investigate how a limiting
procedure might go in the important case of an open event $E\in\bigvee\S$.
In so doing,
let us employ for $E$ the canonical 
approximants $E_n\in\S_n$ ``of type C1'',
these being the simplest to work with and probably the first to suggest
themselves for most people:
$$
    E_n = \bigcup\SetOf{Z\in\Z_n}{Z\part E} \ . 
   \eqno(10)=(7)
$$
As we know, there is no guarantee in general that the corresponding vectors
$\ketvector{E_n}$ will converge, but when they do, we'd like to regard $E$
as ``measurable'' and to associate to it the measure 
$\ketvector{E}=\lim_n\ketvector{E_n}$.   
Below I will illustrate this procedure with the two-site and three-site
hoppers, but first let us consider whether or not our criterion of
convergence is adequate as it stands or whether it needs to be
strengthened. 

Recall in this connection that we had defined a second sequence of
approximants for $E$ ``of type C2'', related to the first ones by
past-completion of the corresponding node-sets $\alfa$.  Would these
approximants have led to the same set of measurable open events and to
the same measures for them?  
A second question concerns compatibility with the Boolean operations,
some form of which is needed if the extended measure is to be
additive on disjoint events.
Consider for example the disjoint union $G=A+B$ of two open events $A$ and
$B$, and let $G_n$, $A_n$ and $B_n$ be the corresponding C1 approximants.
If $\ketvector{A_n}+\ketvector{B_n}=\ketvector{G_n}$ held
automatically it would follow immediately that
$\ketvector{A}+\ketvector{B}=\ketvector{G}$, as desired.
But 
plainly this is not automatic because $G=A\cup B$ can include cylinder
sets $Z$ that are not included separately in either $A$ or $B$
(see figure \Fsita).
We'd like the contribution from such $Z$ to go away in the limit
$n\to\infty$, and we'd also like any  mismatch between our C1 and C2
sequences to go away.
These two desiderata turn out to be closely related.

\vbox{
  \bigskip
  \includegraphics[scale=0.4]{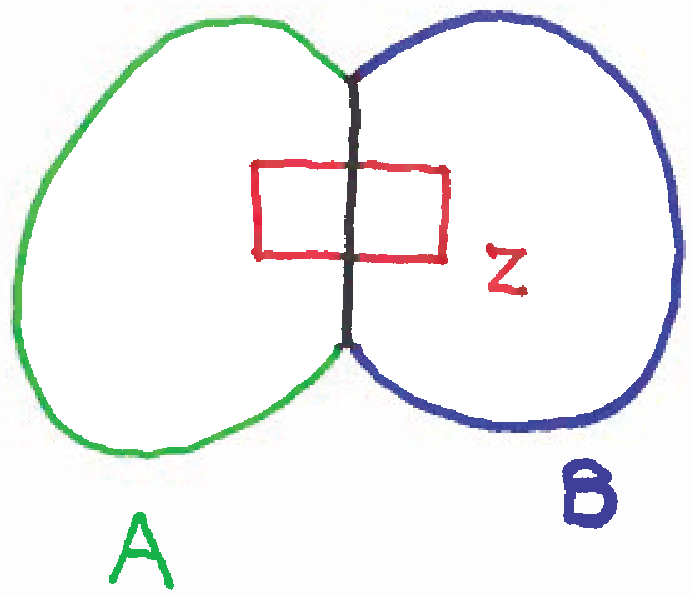}

  \Caption{{\it Figure \Fsita.} A cylinder set $Z$ contributing to the difference (11). }}

Let us examine the difference,
$$
  \ketvector{G_n} - \ketvector{A_n + B_n}
  =
  \ketvector{G_n} - \ketvector{A_n} - \ketvector{B_n} \ ,
  \eqno(11)
$$
more closely.  In light of equation (10), this
difference is just
$$
  \sum\SetOf{\ketvector{Z}}{(Z\in A+B)(Z\notin A)(Z\notin B)} \ ,
$$ 
but the $Z$ here are not arbitrary cylinder sets.  
Rather, I claim that
any $Z$ which contributes to the above sum is
special in that its overlaps with $A$ and $B$ are clopen
(which implies that --- within $Z$ --- the discrepancy 
disappears entirely in a later approximation).
By the next lemma, something similar holds for the difference between
the C1 and C2 approximants to any individual open set $A$.

\DEFINITION The cylinder set $Z$ {\it straddles} the event $A$ if
  it meets both $A$ and its complement and if $Z{\cap}A$ is clopen.
  In symbols: $0\not=ZA\not=Z$ and $ZA\in\S$. 

\REMARK $Z$ straddles $A$ iff it straddles the complement $\Oga\less{}A$

\noindent
To see why the $Z$ contributing to equation (11) are ``straddlers'' in
this sense, it suffices to observe first that $ZA\ideq Z\cap A$ is open
because both $A$ and $Z$ are open, and second that $ZB$ is therefore closed, 
having the form closed$\less$open: $ZB=Z\less ZA$~.  By symmetry both
$ZA$ and $ZB$ are consequently both open and both closed.

\LEMMA  Let $A$ be an open event and let $\alfa$ be the corresponding
  node-set of type C1.  If  past-completion adds $Z\in\Z$ to $\alfa$ 
  then 
  $Z$ straddles $A$, and conversely. 

\PROOF 
Since $A$ is open, $\alfa$ is upward-closed, while for any $\alfa$ at
all, it's true that $\beta=\SetOf{x\in\T}{x\prec\alfa}$ is downward
closed.  Hence the difference $\beta\less\alfa$, the exclusive past of
$\alfa$, is also downward closed; i.e. it is a subtree of $\T$. 
Now suppose without loss of generality that $Z\in\beta\less\alfa$.  By
definition, past-completion will adjoin $Z$ to $\alfa$ iff no path
originating at $Z$ can remain within the subtree $\beta\less\alfa$.  (It
cannot leave $\beta\less\alfa$ and later return to it in this case
because subtrees are convex.)  But this means that the portion of
$\beta\less\alfa$ above $Z$ is actually finite (by the infinity lemma). 
Consequently, for $n$ sufficiently big, every descendant of $Z$ is
either in $\alfa$ or below no node of $\alfa$.
Translated into the language of subsets of $\Oga$, this
says that at a sufficient degree of refinement $n$, every cylinder set
$Z'\in\Z_n$ and within $Z$ is either fully included within $A$ (the
former alternative) or disjoint from $A$ (the latter alternative).
Now suppose that past-completing $\alfa$ does adjoin $Z$ to it.
Then $ZA$ is the union of the $Z'$ belonging to the first family and is therefore
clopen, straddling $A$.
Conversely, if $ZA$ is clopen, then it is a union of cylinder
sets $Z'\in\Z_n$ for some $n$.  \qed

In view of these results, we can to some extent 
deal with the issues raised above
by 
provisionally
adding to our
criterion of convergence the requirement that any ``straddling''
cylinder sets contribute negligibly in measure as $n\to\infty$.

\DEFINITION 
  Let $A_n\in\S_n$ be a sequence of clopen events.
  Call this sequence {\it \evenlyconvergent} with respect to $\mu$ if the following hold:
\item{(i)}  $A = \lim A_n$ for some $A\part\Oga$
\item{(ii)} $\ketvector{A} = \lim \ketvector{A_n}$ for some $\ketvector{A}\in\H$
\item{(iii)} $(\forall \eps > 0)(\exists N)(\forall n > N) \; 
             \sum  || \; \ketvector{Z} \, || < \eps$~,
  where the sum ranges over all $Z\in\Z_n$ that straddle $A$.


\noindent
In view of the previous lemma, condition (iii) implies immediately that
the C1 and C2 approximants for any open event $E$ yield equivalent
results.  As desired, it also gives rise to additivity on disjoint open
events, as we will see in the next theorem.
In the statement of the theorem, the canonical approximants may for
definiteness taken to be those ``of type C1''.  As we have just seen,
exchanging any of them for type C2 would have no effect.
(Notice also in the statement of the theorem that 
the Boolean sum $A+B$ coincides with the union $A\cup B$ 
when $A$ and $B$ are disjoint: $AB=0$.)

\THEOREM  
  Let $A$ and $B$ be disjoint open events and let 
  $A_n$ and $B_n$ be their canonical approximating sequences,
  with $G_n$ being the canonical approximating sequence for $G=A+B$.
  If the first two sequences are \evenlyconvergent then the third is also, and the
  measures add: \hbox{ $\ketvector{A}+\ketvector{B}=\ketvector{G}$ }.

\PROOF  It will be convenient in the following to work with the canonical sequences
  given by (10), since for them the approximants to $A$ and $B$
  will be disjoint.  In the above definition of being \evenlyconvergent, we need to
  establish conditions (i)--(iii) with $A$ replaced by $G$.  
  To begin with,
  condition (i), viz. $G=\lim{G_n}$, is true by construction.  

 \noindent 
  Turning
  to condition (ii), we must verify that
  ~$\ketvector{G}=\lim\ketvector{G_n}$~ with ~$\ketvector{G}=\ketvector{A}+\ketvector{B}$~.
 We've already learned in connection with equation (11) that 
 $\kv{G_n}-\kv{A_n}-\kv{B_n}$ is the sum of the measures $\kv{Z}$ of all
  those cylinder sets $Z\in\Z_n$ that straddle both $A$ and $B$.  But
  this sum can be made arbitrarily small by choosing $n$ big 
  enough, since by hypothesis the sequence $A_n$ is itself \evenlyconvergent. 
  (In more detail:
  $|| \; \kv{G_n}-\kv{A_n}-\kv{B_n} \,|| 
   =
   || \sum \SetOf{\kv{Z}} {Z \hbox{ straddles both } A \hbox{ and } B} ||
  \le
      \sum \SetOf{|| \; \kv{Z} \, ||} {Z \hbox{ straddles both } A \hbox{ and } B} ||  
  \le
      \sum \SetOf{|| \; \kv{Z} \, ||} {Z \hbox{ straddles } A} ||
  \to 0$ as $n\to\infty$.)  
  Therefore
  ~$\lim\kv{G_n}=\lim(\kv{A_n}+\kv{B_n})=\lim\kv{A_n}+\lim\kv{B_n}=\kv{A}+\kv{B}$, 
  as required.

\noindent
  Finally, we need to verify that the $G_n$ themselves fulfill the third
  condition for being \evenlyconvergent.  To that end we will demonstrate that any
  cylinder event $Z\in\Z_n$ that straddles $G$ also straddles either $A$ or $B$.
  The total norm of the straddlers of $G$ will thus be bounded by the
  sum of the bounds for $A$ and $B$, both of which go to zero as $n$
  goes to $\infty$;
  and therewith the proof will be complete.
  Suppose then that $Z$ straddles $G=A+B=A\sqcup B$,
  where the symbol `$\sqcup$' denotes the union of disjoint sets. 
  We have then $ZG = Z(A \sqcup B) = ZA \sqcup ZB$. 
  By definition $Z$ meets $A+B$, so suppose it meets $A$: $ZA\not=0$. 
  Now $ZA$ is obviously open since both $Z$ and $A$ are open.  It is
  also closed, being the difference of the clopen set $ZG=ZA\sqcup ZB$
  and the open set $ZB$.  Hence $Z$ straddles $A$ if it meets $A$ at
  all, and in general it will straddle either $A$ or $B$, as announced. 
  \qed

The theorem takes a first step toward arranging additivity of the
extended measure, but of course disjoint open events is a special
case.  More generally, one would like to have similar theorems covering,
say, arbitrary events in $\R\bigvee\S$ (not just open events) and
arbitrary Boolean operations (not just disjoint union).  
For example, it's easy to establish for any open event $E$ that $\kv{E}$
is defined iff $\kv{\Oga\less E}$ is defined, and that then
$\kv{\Oga\less E}+\kv{E}=\kv{\Oga}$.
To what extent such results can be obtained in general remains to be
investigated.

\subsection{Examples}
Our friend, the return-event $R$,
can serve to illustrate
some of the
definitions we have made.  Let us start with the 2-site hopper, in which
case $R'=\Oga\less{R}$, the event of ``non-return'', consists of the single
history, $(0,0,0,\cdots)$~.  As we know, $R$ itself is topologically open, and
correspondingly $\Oga\less{R}$ is closed, as one can see directly from
the fact that it is the limit of a decreasing sequence of clopen events
of the form $R'_n=\cyl(0,0,0,\cdots,0)$, these being our canonical
approximants for $R'$.  In this case no cylinder event in $\Z_n$ straddles $R'_n$
since it itself is a cylinder event.  To check that $\kv{R'}$ is well
defined, then, we have only to check that the sequence $\kv{R'_n}$
converges.  In fact, it converges trivially to 0, since 
~$|| \; \kv{R'_n} \, || = (1/2)^{n/2}$~.  Thus, $\kv{\Oga\less R}=0$~
and non-return is {\it precluded} for the two-site hopper [14].
Taking complements shows then that $\kv{R}$ is defined and has the value
$\kv{R}=\kv{\Omega}$~.

In the context of the three-site hopper, the events of return and
non-return become much more interesting.
%
Classically, non-return ``almost surely'' does not occur in a finite
lattice; its measure vanishes.  Moreover, this conclusion obtains
independently of what initial conditions one cares to assume.  What we
will find quantally?  More generally, what will we find if, instead of
asking whether the particle visits site $0$, we ask whether it visits
site $1$ or $2$?  By symmetry, these questions become equivalent if we
generalize our initial condition to admit different starting sites.  Let
us therefore consider (still more generally) an initial condition, in
which each possible initial location contributes its own complex
amplitude  $\psi_0(j)$~, $j=0,1,2\in\Integers_3$~.  
%
The measure of a cylinder-set of trajectories can then be derived from
the 3-site analog of equation (2), generalized to allow
for an arbitrary initial position $x_0$~, and with an additional factor
of the initial amplitude ~$\psi_0(x_0)$~ thrown in:
$$
  v_y = (U^{-n})_{y x_n} U_{x_n x_{n-1}}\cdots U_{x_2 x_1} U_{x_1 x_0} \psi_0(x_0)\ .
$$

For the event of non-return, one must sum this expression over all
trajectories $x_j$ such that $x_j\not=0$ for all $j>0$.  Evidently the
resulting vector of components $v_y$ is then given by a matrix product,
of the form 
$(U^{-n}) V^n \psi_0$, where the matrix $V$ is nothing but the matrix $U$
with its first row set to zero.  We can also set the first column of $V$
to zero if we re-express $v$ as
$(U^{-n}) V^{n-1} \psi_1$~,
wherein
$\psi_1$ is just $U\psi_0$ with its first entry set to zero.
This way, $V$ becomes effectively a $2\times2$ matrix.

Recall now that for three sites we have (with $\oga=1^{1/3}$)
$$
   U =
   {1 \over \sqrt{3}}
        \pmatrix{1 & \oga & \oga \cr 
                 \oga & 1 & \oga \cr  
                 \oga & \oga & 1 \cr} 
$$
whence also
$$
   V =
   {1 \over \sqrt{3}}
        \pmatrix{0 & 0 & 0 \cr 
                 0 & 1 & \oga \cr  
                 0 & \oga & 1 \cr} 
$$
For these matrices, powers of $U$ and $V$ can both be evaluated
in essentially the same manner, by writing $U$ or $V$ as a linear
combination of orthogonal projectors.  
Taking $U$ as exemplar, we obtain by adding and subtracting a multiple
of the identity matrix to $U$:
$$
    U = \lambda (1-P) + \sigma P  \ ,
$$
where 
$$
   P = {1 \over 3}
        \pmatrix{1 & 1 & 1 \cr 
                 1 & 1 & 1 \cr  
                 1 & 1 & 1 \cr  }
 \quad
 \hbox{ and }
 \quad
   1-P = {1 \over 3}
        \pmatrix{2 & -1 & -1 \cr 
                 -1 & 2 & -1 \cr  
                 -1 & -1 & 2 \cr  }  \ ,
$$
with
$\lambda = (1-\oga)/\sqrt{3}$ and $\sigma = (1+2\oga)/\sqrt{3}$.
In the same way, defining $Q=1/2 \pmatrix{1&1\cr 1&1\cr}$
(or more correctly as the $3\times3$ matrix with this as its lower right
hand corner), we can obtain $V$ in the form,
$$
    V = \lambda (1-Q) + \rho Q  \ ,
$$
where 
$$
   Q = {1 \over 2}
        \pmatrix{0 & 0 & 0 \cr 
                 0 & 1 & 1 \cr  
                 0 & 1 & 1 \cr  }
 \quad
 \hbox{ and }
 \quad
   1-Q = {1 \over 2}
        \pmatrix{0 &  0 &  0 \cr 
                 0 &  1 & -1 \cr  
                 0 & -1 &  1 \cr  }  \ ,
$$
with $\rho=-\oga^2/\sqrt{3}$.
It follows immediately that
$U^n = \lambda^n (1-P) + \sigma^n P$
and
$V^{n-1} = \lambda^{n-1} (1-Q) + \rho^{n-1} Q$.
Noticing now that $|\rho| = 1/\sqrt{3} < 1$, 
while $|\lambda|=|\sigma|=1$,
we see that in the limit $n\to\infty$
we can drop the second term in $V$, 
without affecting $\kv{R'}$, i.e. without affecting whether the
sequence of approximations $\kv{R'_n}$ converges or what it converges to.
And noticing further that $P(1-Q)=0$, we see that we can also drop that
term in the product $U^{-n}V^{n-1}$, leaving the simple asymptotic form,
$U^{-n}V^{n-1}\sim \lambda^{-n} (1-P) \lambda^{n-1}(1-Q) = (1/\lambda)(1-P)(1-Q)$~.
We thus obtain, modulo an exponentially small correction,
$$
          \kv{R'_n} = {1\over\lambda} (1-P) (1-Q) \psi_1  \ .
$$

This formula leads to a somewhat odd conclusion.  With our original
initial condition that the particle begins at 0, 
the components of $\psi_1$ are just the
last two entries of the first column of $U$, namely
$(\oga/\sqrt{3})(0,1,1)$, which evidently belongs to the kernel of $1-Q$.  
Hence the event of non-return is again precluded: $\kv{\nonreturn}=0$;
and once again $\kv{\return}=\kv{\Oga}=(1,0,0)$.  At first sight, this result
might appear to confirm one's classical intuition, but in fact it seems to
be a coincidence, at least if we take the
3-site hopper as typical.  For almost any other choice of initial
amplitudes
than $(1,0,0)$, 
$\kv{\nonreturn}$ does not vanish!  In particular, if the
particle starts at site 2 instead of site 0, then the event that it
fails to visit site 0 has the non-zero vector-measure
$\kv{\nonreturn}=(1/3,-1/6,-1/6)$. 
(The quantal measure of this same event in the sense of [21] is
$\langle R'|R'\rangle$, or $1/6$.)
The vector-measure of the complementary event that the
particle does visit 0 is then 
$\kv{\return}=\kv{\Oga}-\kv{\nonreturn}=(-1/3,1/6,7/6)$~.


Our analysis of the 3-site case used tacitly the fact that the events
$R$ and $R'$ are free of straddling cylinder-sets, for the same reason
that stem-events are.  Convenient though this is, it means that our
example fails to illustrate condition (iii) in our 
definition of an \evenlyconvergent sequence.  
It would be good to work out an example where (iii) does come
into play, since doing so could indicate whether that condition is a
reasonable one to have added, or whether on the contrary it tends to
rule out events that one would want to include.

It would also be good to work out some physically interesting instances
of our approximation procedure in the causal set case.  One might begin,
for example, with the event ``originary'' for the relatively simple
dynamics of complex percolation.

\section{8.~Epilogue: does physics need actual infinity?}	
Does the description of nature require actual infinities?  Or is a truly
finitary physics possible, in which infinite sets would figure only as
potentialities?


Inasmuch as the theories to which we have grown accustomed employ real
numbers heavily, they thereby presuppose an actual infinity of
cardinality $\alephC$, as emphasized in [22].
In itself, however, this seems more a matter of convenience than of
principle, since one could imagine making do with rational numbers of a
very fine but finite precision that could be made still finer as the
need arose --- in other words a potential infinity.\footnote{$^\star$}
{In writing `$\aleph_1$', I have adopted the continuum hypothesis,
 $\aleph_1=2^{\aleph_0}$, for~\dots~notational reasons.}

The other prominent continuum in present-day physics is of course
spacetime.  Non-relativistically, one could again imagine circumventing
the actual infinities that continuous space and time seem to imply, but
when it comes to relativistic field theories, the new requirement of
{\it locality} appears to force strict continuity on us.  Perhaps one
could get by with only $\aleph_0$ points, say points with rational
coordinates, but even that would still be an actual infinity.

Quantum gravity raises all these questions anew, of course.  String
theory and loop quantum gravity both presuppose background continua, at
least in their current formulations.  Causal dynamical triangulations
and the ``asymptotic safety'' approach retain locality and presuppose
the same type of continuum as classical gravity, albeit not as
background.

With causal sets, the situation seems more fluid.  On one hand, they
transcend locality, but on the other hand they still maintain
{\it{}covariance} in the sense of label-invariance, and that brings with
it an ``infrared'' infinity, as discussed earlier.  An important new
feature, however, is that now the infinity is in some sense pure gauge:
we need it only because we have introduced both an auxiliary time
parameter and a space of ``completed causets'' in order to give a
precise meaning to the concept of sequential growth.  Could it be that a
manifestly covariant formulation of growth dynamics could dispense with
this ``last remaining infinity''?  
Limited to measure theoretic tools inherited from the classical theory
of stochastic processes, we apparently lack the technical means to ask
the question properly.
As things stand, we can
acknowledge at a minimum that being able to refer to completed causets
is very convenient even if it ultimately turns out not to be physically
necessary.  (One can also comment here that the cardinality of a
completed causet, though not finite, is reduced to that of the integers.
On the other hand, the associated sample-space $\Omega$ still has the
cardinality of the continuum.)

Based on this evidence, one could perhaps agree that physics is tending
toward more finitary conceptions, even if it hasn't genuinely reached
them yet.  In particular, even if causal sets are implicitly free of
actual infinities, the available mathematical tools don't let us express
this fact clearly.
Might some of the tools that we seem to lack arise naturally in the
course of attempts, like those above, to extract well-defined
generalized measures from quantal path-integrals and path-sums?

\vbox{
\section{Appendix. Some symbols used, in approximate order of appearance}

\def\parskipsave{\parskip}
\parskip=3pt

$\Omega$ (the sample-space or space of histories),\ \  $\Omega^{physical}$,\ \ $\Omega^{gauge}$,\ \ $\Omega(n)$

$0\part\Oga$ (the empty subset)

$\cyl(c)$ (the cylinder event corresponding to the truncated history $c$)

$\Z$ (the semiring of cylinder events), \ \ $\Z_n$

$\S=\R\Z$ (the Boolean algebra generated by $\Z$ = the finite unions of cylinder sets)

$\S_n$

$\T$  (the tree of truncated histories), \ \ $\T_n$

${\bf 1}^{z}\ideq\exp{2\pi i z}$

$\bigvee\S$, \ \ $\bigwedge\S$, \ \ $\bigvee\bigwedge\S$

$S(\alfa)$, \ \ $\Stilde(\alfa)$ 

$\lim$, \ \ $\Lim$, \ \ $\liminf$, \ \ $\limsup$

$\prec$

$\P$, \ \ $\Pbar$

$A\sqcup B$

$\ketvector{Z}=\mu(Z)$

\parskip=\parskipsave
}


\bigskip
\noindent
I would like to thank Sumati Surya for numerous corrections and/or
suggestions for improving the clarity of the manuscript.  
Research at Perimeter Institute for Theoretical Physics is supported in
part by the Government of Canada through NSERC and by the Province of
Ontario through MRI.

\ReferencesBegin                             



\ref [1] A.N.~Kolmogorov and S.V.~Fomin, 
   {\it Measure, Lebesgue Integrals, and Hilbert Space}
   translated by Na\-tascha Artin Brunswick and Alan Jeffrey (Academic Press 1961)

\ref [2] Robert Geroch, ``Path Integrals'' (unpublished notes)
    available at  \lbr
    http://www.perimeterinstitute.ca/personal/rsorkin/lecture.notes/geroch.ps


\ref [3] Fay Dowker, Steven Johnston, Sumati Surya, ``On extending the Quantum Measure'',
   \eprint{arXiv:1007.2725 [gr-qc]}

\ref [4] Graham Brightwell, Fay Dowker, Raquel S.~Garc{\'\i}a, Joe Henson and Rafael D.~Sorkin,
``{$\,$}`Observables' in Causal Set Cosmology'',
\journaldata {Phys. Rev.~D} {67} {084031} {2003}
\lbr
\arxiv{gr-qc/0210061}
\lbr
\eprint{http://www.perimeterinstitute.ca/personal/rsorkin/some.papers/}

\ref [5] Graham Brightwell, {H. Fay Dowker}, {Raquel S. Garc{\'\i}a}, {Joe Henson} and {Rafael D.~Sorkin},
``General Covariance and the `Problem of Time' in a Discrete Cosmology'',
 in K.G.~Bowden, Ed., 	
 {\it Correlations}, 
 Proceedings of the ANPA 23 conference,
 held August 16-21, 2001, Cambridge, England 
 (Alternative Natural Philosophy Association, London, 2002), pp 1-17
\eprint{gr-qc/0202097}
\lbr
\eprint{http://www.perimeterinstitute.ca/personal/rsorkin/some.papers/}

\ref [6] Luca Bombelli, Joohan Lee, David Meyer and Rafael D.~Sorkin, 
``Spacetime as a Causal Set'', 
  \journaldata {Phys. Rev. Lett.}{59}{521-524}{1987}

\ref [7] Rafael D.~Sorkin, ``Causal Sets: Discrete Gravity (Notes for the Valdivia Summer School)'',
in {\it Lectures on Quantum Gravity}
(Series of the Centro De Estudios Cient{\'\i}ficos),
proceedings of the Valdivia Summer School, 
held January 2002 in Valdivia, Chile, 
edited by Andr{\'e}s Gomberoff and Don Marolf 
(Plenum, 2005)
\lbr
\arxiv{gr-qc/0309009}
\lbr
\eprint{http://www.perimeterinstitute.ca/personal/rsorkin/some.papers/} 

\ref [8] Joe Henson, ``The causal set approach to quantum gravity''
 \eprint{gr-qc/0601121}
 This is an extended version of a review to be published in 
 in  {\it Approaches to Quantum Gravity -- Towards a new understanding of space and time}, 
 edited by Daniele Oriti 
 (Cambridge University Press 2009)
 (ISBN: 978-0-521-86045-1), pages 26-43,

\ref[10] Sumati Surya,  ``Directions in Causal Set Quantum Gravity'',
\eprint{arXiv:1103.6272v1 [gr-qc]}

\ref [9] Fay Dowker, ``Causal sets and the deep structure of Spacetime'', 
 in
 {\it 100 Years of Relativity - Space-time Structure: Einstein and Beyond}"
 ed Abhay Ashtekar 
 (World Scientific, to appear)
 \eprint{gr-qc/0508109}

\ref [11] David P.~Rideout and Rafael D.~Sorkin,
``A Classical Sequential Growth Dynamics for Causal Sets'',
 \journaldata{Phys. Rev.~D}{61}{024002}{2000}
 \eprint{gr-qc/9904062}
 \lbr
 \eprint{http://www.perimeterinstitute.ca/personal/rsorkin/some.papers/}

\ref [12] Alexander S. Kechris, {\it Classical Descriptive Set Theory} (Springer-Verlag 1995)

\ref [13] Yiannis N. Moschovakis, {\it Descriptive Set Theory}, second edition 
 (American Mathematical Society 2009)

\ref [14] S.~Gudder and Rafael D.~Sorkin, ``Two-site quantum random walk'' 
  (in preparation)

\ref [15] Philip Pearle, ``Finite-Dimensional Path-Summation Formulation for Quantum Mechanics''
\journaldata{Phys. Rev. D}{8}{2503-2510}{1973}


\ref [16] Xavier Martin, Denjoe O'Connor and Rafael D.~Sorkin,
``The Random Walk in Generalized Quantum Theory''
\journaldata {Phys. Rev.~D} {71} {024029} {2005}
\eprint{gr-qc/0403085}
\lbr
\eprint{http://www.perimeterinstitute.ca/personal/rsorkin/some.papers/}

\ref [17] Rafael D.~Sorkin, ``Logic is to the quantum as geometry is to gravity''
in G.F.R. Ellis, J. Murugan and A. Weltman (eds),
{\it Foundations of Space and Time} 
(Cambridge University Press, to appear)
\arxiv{arXiv:1004.1226 [quant-ph]}
\eprint{http://www.perimeterinstitute.ca/personal/rsorkin/some.papers/} 
 %


\ref [18] Madhavan Varadarajan and David Rideout,		
``A general solution for classical sequential growth dynamics of Causal Sets''
\journaldata {Phys. Rev. D} {73} {104021} {2006}
\eprint{gr-qc/0504066}




\ref [19] Rafael D.~Sorkin, ``Quantum Measure Theory and its Interpretation'', 
  in
   {\it Quantum Classical Correspondence:  Proceedings of the $4^{\rm th}$ Drexel Symposium on Quantum Nonintegrability},
     held Philadelphia, September 8-11, 1994,
    edited by D.H.~Feng and B-L~Hu, 
    pages 229--251
    (International Press, Cambridge Mass. 1997)
    \eprint{gr-qc/9507057}
   \eprint{http://www.perimeterinstitute.ca/personal/rsorkin/some.papers/}

\ref [20] Rafael D. Sorkin, ``Quantum dynamics without the wave function''
 \journaldata{J. Phys. A: Math. Theor.}{40}{3207-3221}{2007}
 (http://stacks.iop.org/1751-8121/40/3207)
 \eprint{quant-ph/0610204} 
 \lbr
 \eprint{http://www.perimeterinstitute.ca/personal/rsorkin/some.papers/}


\ref [21] Rafael D.~Sorkin, ``Quantum Mechanics as Quantum Measure Theory'',
   \journaldata{Mod. Phys. Lett.~A}{9 {\rm (No.~33)}}{3119-3127}{1994}
   \eprint{gr-qc/9401003}
   \lbr
   \eprint{http://www.perimeterinstitute.ca/personal/rsorkin/some.papers/}

\ref [22] C.J.~Isham, ``Some Reflections on the 
   Status of Conventional Quantum Theory when Applied to Quantum Gravity''  
 \arxiv{quant-ph/0206090v1}

\end                                         


(prog1 'now-outlining
  (Outline* 
     "\f"               1
      "
      "
      "
      "\\Abstract"      1
      "\\section"       1
      "\\subsection"    3
      "\\appendix"      2
      "\\Reference"     1
      "\\ref "          4
      "\\end